\documentclass[eqsecnum,aps,prb,twocolumn,epsfig]{revtex4}

\def\h{\hat{h}}
\def\w{\tilde{w}}
\def\G{\tilde{G}}
\def\H{{\cal{H}}}
\def\W{{\cal{W}}}
\def\a{\overline{a}}
\def\K{\overline{K}}
\def\B{\overline{B}}
\def\v{\overline{v}}

\def\n{\hat{n}}
\def\z{\hat{z}}
\def\E{\hat{E}}
\def\L{\hat{L}}
\def\Hh{\hat{\cal H}}
\def\E{\hat{E}}
\def\q{{\bf q}}
\def\r{{\bf r}}
\def\x{{\bf x}}
\def\zz{{\bf z}}
\def\Hf{{\bf H}}

\begin{document}
\tighten
\title{Thermal depinning and transverse-field tilting transitions in 
a planar vortex array pinned by a columnar defect}
\author{Leo Radzihovsky}
\affiliation{Department of Physics, University of Colorado, Boulder, CO 80309}
\date{\today}
\begin{abstract}
  We study thermal and a transverse magnetic field response of a
  vortex line array confined to a plane with a single columnar pinning
  defect.  By integrating out ``bulk'' degrees of freedom away from
  the columnar defect we reduce this two-dimensional problem to a
  one-dimensional one, localized on the defect and exhibiting a
  long-range elasticity along the defect.  We show that as a function
  of temperature, for a magnetic field aligned with the defect this
  system exhibits a one-dimensional analog of a roughening transition,
  with a low-temperature ``smooth'' phase corresponding to a vortex
  array pinned by the defect, and a high-temperature ``rough'' phase
  in which at long scales thermal fluctuations effectively average
  away pinning by the defect.  We also find that in the
  low-temperature pinned phase, the vortex lattice tilt response to a
  transverse magnetic field proceeds via a soliton proliferation
  ``transition'', governed by an integrable sine-Hilbert equation and
  analogous to the well-known commensurate-incommensurate transition
  in sine-Gordon systems.  The distinguishing feature here is the
  long-range nature of the one-dimensional elasticity, leading to a
  logarithmic soliton energy and interaction. We predict the
  transverse-field---temperature phase diagram and discuss extension
  of our results to a bulk vortex array in the presence of a dilute
  concentration of columnar defects.
\end{abstract}


\maketitle

%\begin{multicols}{2}

\section{Introduction}
\label{introduction}

\subsection{Background and motivation}
\label{background}

The discovery of high-temperature superconductors almost 20 years ago,
among other things, has rekindled interest in the magnetic field ($H$)
- temperature ($T$) phase diagram of type II
superconductors\cite{Tinkham}, generating a vigorous scientific
activity. As a result, much has been clarified about the nature of
equilibrium and nonequilibrium properties of vortex states in the
presence of thermal fluctuations, pinning disorder and electrical
(``super''-) current\cite{BlatterRMP,NattermannScheidl,FFH,HR,driven},
leading to a rich phase diagram. In particular, in contrast to a
mean-field phase diagram, thermal fluctuations drive a first-order
melting of a vortex lattice over a large portion of the phase diagram
into a resistive (although with large conductivity and diamagnetic
response) vortex
liquid\cite{Eilenberger,DSFisher,NelsonSeung,meltingExp}, that, from
the symmetry point of view is qualitatively identical to the normal
state. Furthermore, as was first shown by Larkin\cite{Larkin,ImryMa},
arbitrarily weak pinning disorder, on sufficiently long Larkin scale
(that diverges in the limit of vanishing disorder) always disrupts
translational order of the vortex lattice. It has been argued
theoretically\cite{MatthewFisher,FFH}, with a limited experimental
support\cite{Koch}, that in the resulting state, vortices are
collectively pinned into a vortex glass characterized by an
Edwards-Anderson\cite{EA} like order parameter, exhibiting a vanishing
linear mobility and therefore a vanishing linear resistivity.

While the original proposal for the vortex glass state was made in a
context of intrinsic, short-range correlated (point) disorder, it was
soon appreciated that the beneficiary effects of pinning can be
enhanced by introducing artificial pinning centers by, for example,
electron and/or heavy ion irradiation, with the latter resulting in a
forest of columnar pinning defects, that is a particularly effective
pinning mechanism\cite{columnarExp}. The resulting {\em anisotropic} vortex
glass was dubbed a Bose-glass\cite{FisherLee,NelsonVinokur} because of
its mathematical connection with interacting two-dimensional (2D)
quantum bosons pinned by a quenched (time-independent) random 2D
potential\cite{BoseGlass}.  This connection allowed understanding many
of the properties of the anisotropic vortex glass from the
corresponding quantum Bose-glass phase.\cite{BoseGlass}

One key feature of the anisotropic vortex glass that distinguishes it
from the corresponding (putative) isotropic one is the existence of
the ``transverse'' Meissner effect,\cite{NelsonVinokur} namely a
vanishing response to a field $H_\perp < H_\perp^c$ applied
transversely to columnar defects and vortex lines.  This expulsion of
the transverse flux density, $B_\perp$, that has received considerable
experimental\cite{trMeissnerExp} and simulations\cite{Wallin} support,
corresponds to an effectively divergent anisotropic vortex glass tilt
modulus\cite{BalentsEPL}, that in the quantum correspondence maps onto
a vanishing superfluid density in the Bose-glass phase. The detailed
theoretical description of the transverse Meissner effect (as well as
other properties of the phase) has been predominantly limited to
noninteracting vortex lines\cite{NelsonVinokur,Hwa}. Although these
are supported by scaling
theories\cite{NelsonVinokur,Hwa,NelsonRadzihovsky,Wallin} (borrowed
from the variable-range hopping theory for electronic
systems\cite{EfrosShklovskii}) that do incorporate effects of both
disorder and interactions (clearly essential for the very existence of
the Bose-glass phase), with the exception of functional RG
analysis\cite{BalentsEPL}, a detailed interacting description is
limited to simulations.\cite{Wallin,TauberNelson} This is not
surprising, as a description of strongly interacting random systems is
a notoriously difficult (with few exceptions) unsolved problem, whose
solution is at the heart of understanding many of the interesting
condensed matter phenomena.

One way to incorporate strong interactions is to approach the problem
from the vortex solid (rather than the vortex liquid starting point)
pinned by a random potential. Potential difficulties with this
approach are a proper incorporation of topological defects
(dislocations and disclinations) that tend to proliferate in the
presence of quenched disorder and external perturbations.

Recent analytical real-space renormalization group (RG)
study\cite{DSFisherBG} that demonstrated stability of a 3D weakly
disordered random-field XY model to a proliferation of topological
defects (vortices), provide a strong argument for the stability of an
elastically disordered but topologically ordered vortex Bragg-glass
phase postulated and studied in detail by Giamarchi and Le
Doussal\cite{GL}. These studies therefore give support to treatments
of vortex
solids\cite{NattermannBG,Korshunov,DSFisherFRG,MatthewFisher} that
ignore the notoriously difficult-to-treat topological
defects. Furthermore, even if Bragg-glass {\em is} unstable to
dislocations, for weak disorder dislocations will be dilute, with
physics on scales smaller than their spacing expected to be
well-described by the vortex Bragg-glass
phenomenology.\cite{NattermannBG,Korshunov,DSFisherFRG,MatthewFisher,GL}

A suppression of topological defects can furthermore be facilitated by
a planar confinement of vortices, realized in layered high-T$_c$
superconductors (e.g., BISCCO)\cite{Kwok} or by artificially prepared
multilayers\cite{Kapitulnik,Moshchalkov}, where for a magnetic field
directed along the planes, vortices are well localized to 2D.  The
resulting planar (2D) vortex array pinned by point disorder, where
dislocations are excluded by construction is in fact the ``toy'' model
studied by Matthew Fisher\cite{MatthewFisher}, that motivated his
original proposal of a vortex glass phase in bulk superconductors.

\begin{figure}[bth]
\centering
\setlength{\unitlength}{1mm}
\begin{picture}(40,50)(0,0)
\put(-23,2){\begin{picture}(30,40)(0,0)
\includegraphics{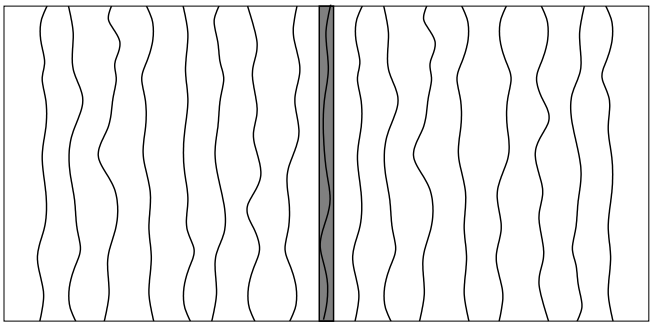}
\end{picture}}
%\put(-20,40) {$n_p$}
%\put(60,2) {$\epsilon_p$}
\end{picture}
\caption{A (1+1)-dimensional (planar) vortex lattice pinned by a single 
columnar defect studied in this paper.}
\label{2Dlattice}
\end{figure}

Motivated by these ingredients, in this paper we study a
(1+1)-dimensional vortex array confined to a planar slab of thickness
$w$ in a presence of a single planar columnar defect, illustrated in
Fig.\ref{2Dlattice}.  This system was introduced and first studied in
great detail in Ref.~\onlinecite{Affleck}.

\begin{figure}[bth]
\centering
\setlength{\unitlength}{1mm}
\begin{picture}(40,50)(0,0)
\put(-23,2){\begin{picture}(30,40)(0,0)
\includegraphics{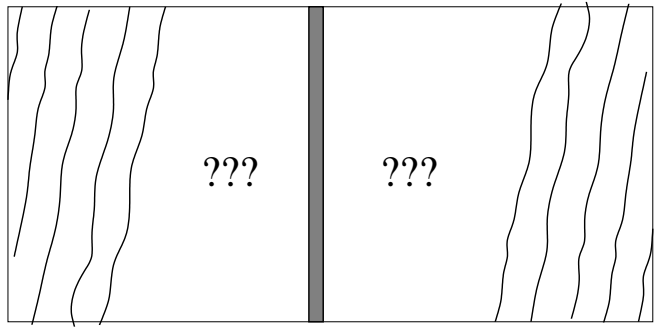}
\end{picture}}
%\put(-20,40) {$n_p$}
%\put(60,2) {$\epsilon_p$}
\end{picture}
\caption{A (1+1)-dimensional (planar) vortex array induced by a magnetic 
field $\Hf$ applied in the plane of the slab, at an angle to a single
columnar pinning defect.}
\label{2DlatticeTilted}
\end{figure}

As should be clear from the above discussion, such a ``toy''
model\cite{Affleck} may be relevant to the regime of far-separated (by
$d$, compared to vortex spacing $\sqrt{\phi_0/B}$) columnar defects,
accessible for flux density far exceeding the columnar-defect matching
field $B_\phi=\phi_0/d^2$,\cite{NelsonVinokur,RadzihovskyPRL} where
$\phi_0=hc/2e\approx 2.1\times10^{-7}$G-cm$^2$ is a fundamental
quantum of flux.  As first investigated in Ref.~\onlinecite{Affleck},
we study the response of such a (1+1)-dimensional vortex array to a
planar tilting magnetic field $H_\perp$, applied transversely to the
columnar defect, as illustrated in Fig.\ref{2DlatticeTilted}.

The rest of the paper is organized as follows. We conclude the
Introduction with a summary of our main results and predictions,
heuristically extended to a dilute concentration of columnar defects.
In Sec.\ref{model} we derive the appropriate (1+1)-dimensional
continuum model for a single defect and discuss its ingredients.  By
integrating out ``bulk'' (away from the defect) degrees of freedom we
reduce this model to a (0+1)-dimensional model confined to a defect,
and characterized by a long-range elasticity along the defect. In
Sec.\ref{Hperp} we study the effect of transverse magnetic field
$H_\perp$ and demonstrate that tilting of the vortex lattice away from
a columnar defect proceeds via a novel ``commensurate-incommensurate
transition'' controlled by a proliferation of solitons. In
Sec.\ref{finiteT} we study effects of thermal fluctuations and
demonstrate that this system exhibits a 1D ``roughening''-like
transition. We explore its consequences for the vortex positional
correlations, and construct a $H_\perp$--$T$ phase diagram.  We
conclude in Sec.\ref{finiteDefects} with an extension of these results
to an experimentally relevant case of a dilute concentration of
columnar defects (allowing for genuine transitions) and close in
Sec.\ref{conclusion}.with a summary of our study.

\subsection{Summary of results}
\label{results}

The body of the paper is primarily devoted to the study a planar
(1+1)-dimensional vortex array at a 1D vortex density $n_0=1/a$, in
the presence of thermal fluctuations and induced by an external planar
magnetic field $\Hf=H_z{\bf \z}+\Hf_\perp$ applied at an angle to a
single columnar defect\cite{Affleck}.  As we will show below, even
this ``toy'' problem is quite rich, providing insight into the bulk
(2+1)-dimensional multi-defect problem.  It has the added benefit that
it can be analyzed in detail analytically. This is not surprising as
such a planar classical vortex array is a cousin of a one-dimensional
quantum problem, a Luttinger liquid, that is known to be exactly
solvable and to exhibit rich phenomenology.\cite{HaldaneLL} In fact
our classical analysis of the vortex problem has strong formal
connections to the work of Kane and Fisher\cite{KaneFisher} who
studied a Luttinger liquid in a presence of a single localized
impurity, a problem that admits exact analysis\cite{Fendley}. This
connection was first emphasized and fruitfully utilized by Hoffstetter
{\em et al.}\cite{Affleck} and Polkovnikov, {\em et
  al.}\cite{Polkovnikov}, although we will not take advantage of it.
As is well known\cite{HaldaneLL}, the Luttinger formalism is
equivalent to the classical theory of vortex lattice elasticity that
we (and Ref.~\onlinecite{Affleck}) employ here.

Our work has a strong overlap with that of Ref.~\onlinecite{Affleck},
particularly on the finite temperature analysis for a vanishing
transverse field in an infinite single-pin system, as well as a large
transverse field, where tilt response is analytic. Where this overlap
exists our predictions are in complete agreement with those found in
Ref.~\onlinecite{Affleck}. However, our emphasis is on the
low-temperature, strong coupling regime, where tilt response is highly
nonlinear, and can only be understood in detail in terms of vortex
lattice solitons, that proliferate at a novel
commensurate-incommensurate crossover, that, we argue, turns into a
genuine sharp phase transition for a dilute concentration of columnar
defects.

\begin{figure}[bth]
\centering
\setlength{\unitlength}{1mm}
\begin{picture}(40,70)(0,0)
\put(-22,-2){\begin{picture}(30,40)(0,0)
\includegraphics{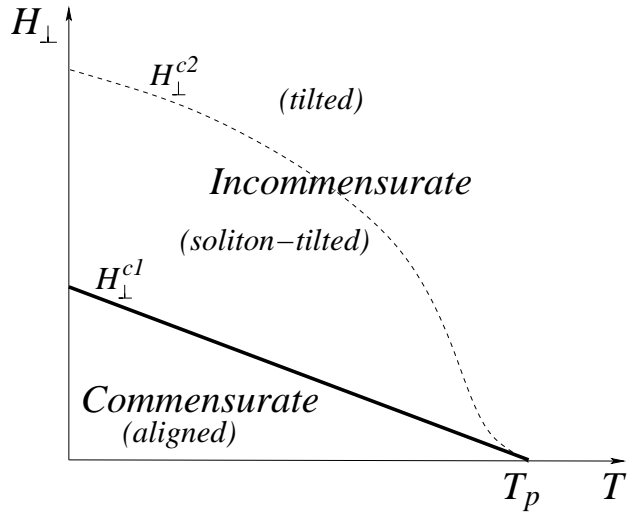}
\end{picture}}
%\put(-20,40) {$n_p$}
%\put(60,2) {$\epsilon_p$}
\end{picture}
\caption{$H_\perp$ -- $T$ phase diagram illustrating the commensurate (C) 
  and incommensurate (I) phases separated by a continuous CI phase
  transition at $H_\perp^{c1}(T)$ (full curve), where vortex lines in
  the vicinity of far-separated columnar defects tilt via
  proliferation of solitons. As the soliton density grows with
  increasing $H_\perp$ (dashed curve), the solitons overlap near the
  upper-critical transverse field $H_\perp^{c2}(T)$, beyond which the
  system crosses over to a smoothly tilted vortex lattice. The
  lower-critical field, $H_\perp^{c1}(T)$ vanishes with increasing pin
  separation $d$, eliminating the CI transition (but not the
  roughenning transition at $T_p$) for a single columnar defect.}

\label{phasediagram}
\end{figure}

Hence, as we describe in more detail below, in the presence of a
dilute concentration of columnar defects the planar vortex array
exhibits two phases in the transverse field $H_\perp$ -- temperature
$T$ phase diagram: a low $T$, $H_\perp$ ``commensurate''
pinned/aligned (C) phase and a high $T$, $H_\perp$ ``incommensurate''
depinned/tilted (I) phase. As illustrated in Fig.\ref{phasediagram}, a
novel finite temperature commensurate-incommensurate phase transition
separates the two phases.

For a vanishing transverse field, the phase transition is akin to a
thermal roughening transition\cite{roughening,Pokrovsky,Coppersmith},
that is closely related to a zero-temperature superfluid-insulator
transition in a resistively-shunted Josephson junction\cite{RSJ}, the
opaque-to-transparent impurity transition in a Luttinger
liquid\cite{KaneFisher}, and spin-boson and dissipative impurity
models\cite{1dIsing,Leggett,Kondo}.

Before we discuss our results, a disclaimer is in order here. Because
a {\em single} columnar defect cannot possibly compete with bulk
degrees of freedom, with its effects vanishing in thermodynamic limit,
we expect a two-dimensional free energy that is {\em analytic}, and
therefore no true phase transition can take place in a 2D
thermodynamic limit. However, as in, for example, the Kondo
problem\cite{Kondo}, where one considers the effect of an impurity on
the bulk electron gas, bulk effects are predicted only once a finite
density of defects is considered.  Here too the transitions that we
discuss are for the elastic degrees of freedom localized on the
columnar defect, that is a {\em boundary} critical phenomenon.
Although the 1D roughening transition is indeed a genuine one, the
contribution of the associated (nonanalytic) free energy to the bulk
two-dimensional system vanishes in the 2D thermodynamic limit.
Furthermore, the CI (vortex line tilting) ``transition'' takes place
at a lower-critical transverse field $H_\perp^{c1}(L)$, that, in the
case of one pin is driven to zero in the thermodynamic limit
($L\rightarrow\infty$).  It only becomes a genuine sharp phase
transition for a finite density $1/d$ of columnar defects, with the
lower-critical transverse field $H_\perp^{c1}(d)$ set by the
columnar-pin spacing $d$.\cite{noTransition} Since this latter case is
the one most easily accessible experimentally, we heuristically extend
our rigorous single pin results to a finite dilute concentration of
independent columnar defects. This being a notoriously difficult
unsolved problem, we expect this heuristic extension to break down on
sufficiently long scales, where collective pinning effects become
important.
\cite{FisherLee,NelsonVinokur,Hwa,BalentsEPL,NelsonRadzihovsky,Wallin}

Our results can be divided into two categories: thermal effects at a
vanishing transverse field, i.e., for the magnetic field (and
therefore induced planar vortex array) aligned with columnar defects,
and a low-temperature response of the vortex array to a tilting
(transverse) magnetic field.

For a magnetic field aligned with columnar defects we show that a
vortex array undergoes a ``roughening''-like transition at
\begin{equation}
T_p={\sqrt{K B}\over \pi n_0^2},
\label{Tc}
\end{equation}
between a low-temperature state in which each columnar defect
effectively pins the vortex lattice around the pin for $T<T_p$ and a
phase in which its pinning effects, even near a columnar defect,
vanish at long scales. In above, $K$ and $B$ are vortex lattice tilt
and compressional moduli, whose vortex density and dependence on other
parameters (e.g., vortex interaction) can be computed from
microscopics\cite{Coppersmith,elasticity,GLelastic,Affleck}. This
roughening-like transition is related to a one-dimensional long-range
interacting ($1/|i-j|^2$) Ising model\cite{1dIsing}, that is
well-known from a number of other physical contexts, most prominently
the Kondo problem\cite{Kondo}, and more recently, in a dissipative
Josephson junction\cite{RSJ} and a Luttinger liquid in the presence of
an impurity\cite{KaneFisher}. In relation to the latter work, we note
that unlike the quantum case where there is no simple way to tune the
Luttinger parameter\cite{FBR}, here it can be simply tuned by
temperature and vortex density.\cite{gless1}

As is usually the case for such (e.g.,
roughening\cite{roughening,Pokrovsky,Coppersmith},
Kosterlitz-Thouless\cite{KT}, and other topological) transitions there
is no local order parameter and phases are distinguished by long-scale
behavior of correlation functions. The low-temperature $T<T_p$, pinned
phase is characterized by a finite mean-squared vortex fluctuations at
the location of the defect (that we take to be $x=0$), with
correlations away from the defect given by\cite{Affleck}
\begin{eqnarray}
\langle u(z,x) u(0,x')\rangle\approx {k_B T\over 4\pi\sqrt{K B}}\ 
\ln\left[{K(|x|+|x'|)^2 + B z^2\over K(x - x')^2 + B z^2}\right].\nonumber\\
\label{uuPinned_results}
\end{eqnarray}

On scales longer the pinning length $\xi$ (defined in Eq.\ref{xi},
below), the corresponding average density exhibits Friedel-like
oscillations\cite{Affleck} given by
\begin{eqnarray}
\langle n(x,z)\rangle_0 
&\approx& n_0 + 2n_G\left({\a\over 2|x|}\right)^{\eta/2}\cos(2\pi n_0 x),
\label{n0pinned_results}
\end{eqnarray}
where $\a={\rm Max}[a,\xi\sqrt{B/K}]$ and
\begin{equation}
\eta={2\pi\over a^2}{k_B T\over\sqrt{K B}}.
\end{equation}

In contrast, at high temperature $T>T_p$, vortex thermal fluctuations
effectively average away the effects of the pin, leading to phonon
correlations that diverge logarithmically with sample size. The
connected phonon correlation function is finite and is given by
\begin{eqnarray}
\langle(u(x,z)-u(0,0))^2\rangle &\approx&
{k_B T\over\pi\sqrt{K B}}\ln\left[a^{-1}\sqrt{x^2+{B\over K}z^2}\right].
\nonumber\\
\label{uuDepinned_results}
\end{eqnarray}

Despite the irrelevance (in the RG sense of the term) of the columnar
defect for $T>T_p$, the average density also displays perturbative
Friedel oscillations\cite{Affleck}
\begin{eqnarray}
\langle n(x,z)\rangle &\approx&n_0 + 
{n_G a v\over k_B T}\sqrt{K\over B}
\left({a\over|x|}\right)^{\eta-1}\cos(2\pi n_0 x),\nonumber\\
&&
\label{n1intro}
\end{eqnarray}
with a stronger power-law exponent than that for $T<T_p$ and an
amplitude that vanishes with the strength of the pinning potential
$v$.

The low-temperature $T<T_p$, the pinned (commensurate) phase is
distinguished from the rough (incommensurate) phase by a transverse
Meissner response to a magnetic field $H_\perp$ applied transversely
to the columnar pin.  Namely, we find that for a field smaller than a
lower-critical transverse field $H^\perp_{c1}$
\begin{mathletters}
\begin{eqnarray}
\hspace{-1cm}
H^\perp_{c1}&\approx&
{\phi_0\over w}\cases{{1\over L}\ln{L\over\xi},
& $L\sqrt{B\over K}\ll d$\cr 
\sqrt{B\over K}{1\over4\pi d}
\ln\left[\sqrt{K\over B}{2\pi d\over\xi}\right],& $L\sqrt{B\over K}\gg d$\cr},
\label{Hc1results}
\end{eqnarray}
\end{mathletters}
where $L$ is the length of the sample along the columnar defect ($z$),
$w$ is the slab thickness, and $d$ is the columnar pin spacing.  At
low temperature $\xi\approx\xi_0$ is approximately given by
\begin{eqnarray}
\xi_0&\approx&{a^2\over 2\pi^2}{\sqrt{B K}\over v}.
\label{xi_results}
\end{eqnarray}
For $H^\perp < H^\perp_{c1}$ vortex lines in the wide vicinity
\begin{equation}
\lambda_h^{0}=\left({B\over K}\right)^{1/2}{L\over\pi}
\label{lambda_0}
\end{equation}
of a columnar pin remain aligned with it, therefore exhibiting a bulk
transverse Meissner effect for a $d < \lambda_h^0$ spaced array of
pins.  Of course, (as for the thermally-driven depinning transition
discussed above, here too), because a single pin cannot compete with
the bulk magnetic energy, away from the defect beyond this screening
length $\lambda_h^{0}$ the vortex lattice is always aligned along the
applied magnetic field. Related to this, for a single pin
($d\rightarrow\infty$) the critical transverse field $H^\perp_{c1}$,
Eq.\ref{Hc1results} clearly vanishes in the thermodynamic $L\to\infty$
limit.

For a transverse field stronger than $H^\perp_{c1}$ a continuous
vortex lattice tilting transition takes place into a tilted
(incommensurate) state. It proceeds via a proliferation of solitons,
with soliton density $n_s(H_\perp)$ and the average vortex tilt at the
location of the columnar defect growing continuously beyond
$H^\perp_{c1}$
\begin{mathletters}
\begin{eqnarray}
n_s(H_\perp)&=&a^{-1}\langle\partial_z u\rangle|_{x=0},\\ &\sim&
\cases{0,& $H < H^\perp_{c1}$\cr |H^\perp - H^\perp_{c1}|, & $H >
H^\perp_{c1}$,\cr}
\label{ns_results}
\end{eqnarray}
\end{mathletters}
In the incommensurate phase, the screening length out to which the
vortex lines are aligned along and pinned by the columnar defect
diminishes with increasing soliton density and is given by
\begin{eqnarray}
\lambda_h&=&\left({B\over K}\right)^{1/2}{1\over2\pi n_s(H_\perp)},\\
&\sim& {1\over|H^\perp - H^\perp_{c1}|}.
\label{lambda_results}
\end{eqnarray}
As we show below, for finite transverse field $H_\perp$, Friedel
oscillations decay {\em exponentially} away from the columnar defect
with length also given by $\lambda_h$.  We estimate the upper-critical
transverse field $H^\perp_{c2}$, for which soliton lattice becomes
dense, to be given by
\begin{eqnarray}
H^\perp_{c2}&\approx& H^\perp_{c1}+{\phi_0\over 2\pi w \xi}.
\label{Hc2results}
\end{eqnarray}
In the presence of fluctuations, $H_{c2}$ field marks a crossover from
a nonlinear soliton tilted regime to a uniformly tilted state. For
large $H_\perp$ exceeding $H^\perp_{c2}$, the vortex lattice tilts
smoothly and the screening length reduces to $\lambda_h\sim
1/H^\perp$, in this limit coinciding with Friedel oscillations decay
length found by Affleck, et al\cite{Affleck}.

The finite temperature CI phase boundary in Fig.\ref{phasediagram} is
given by
\begin{eqnarray}
H^\perp_{c1}(T)&\approx&H^\perp_{c1}(0)\left(1 - {T\over T_p}\right)
\label{Hc1Tresults}
\end{eqnarray}
The corresponding upper-critical crossover field $H^\perp_{c2}(T)$
(dashed curve in Fig.\ref{phasediagram}) for weak pinning ($\xi_0\gg
a\sqrt{K/B}$) can be estimated by using $H_{c1}(T)$ and $\xi(T)$
inside Eq.\ref{Hc2results},
\begin{eqnarray}
H^\perp_{c2}(T)&\approx&H^\perp_{c1}(T) +
H^\perp_{c2}(0)\left({a\sqrt{K/B}\over\xi_0}\right)^{T/|T_p-T|}.
\label{Hc2Tresults}
\end{eqnarray}
For strong pinning ($\xi_0\ll a\sqrt{K/B}$), $H_{c2}^\perp(T)$ is
given by the above expression, but with $\xi_0$ set equal to
$a\sqrt{K/B}$.

In the remainder of the paper we demonstrate results summarized above.

\section{Model}
\label{model}

\subsection{Vortex lattice elasticity with a transverse magnetic field}

In a type-II superconductor, for fields above a lower-critical field
magnetic flux penetrates in a form of interacting vortex flux tubes,
with average density determined by the applied magnetic
field.\cite{Tinkham} At low temperature and in the absence of
disorder, a periodic array (Abrikosov lattice) of repulsive, elastic
vortex lines forms, whose elastic description\cite{elasticity} can be
derived from the Ginzburg-Landau theory for the superconducting order
parameter, that itself, under certain conditions, is derivable from
the microscopic theory of superconductivity.  As usual, transcending
such a detailed derivation, on sufficiently long length scale the
elastic vortex lattice energy functional can be deduced purely on
symmetry grounds. In $d$-dimensions, it is formulated in terms of a
$d-1$-dimensional Eulerian phonon field (Goldstone mode of the
spontaneously broken translational symmetry) ${\bf u}({\bf x},z)={\bf
x}-{\bf x}_i$ describing a transverse vortex lattice distortion at a
$d$-dimensional position $({\bf x},z)$ relative to a perfect vortex
array characterized by $({\bf x}_i,z)$.

A planar vortex array, that we take to be confined to the $x-z$ plane,
is characterized by a scalar phonon field $u(x,z)$, describing
$x$-directed vortex distortion with a continuum elastic Hamiltonian
given by
\begin{equation}
\H_{el}={1\over2}\int dx dz\left[K(\partial_z u - h)^2 + B(\partial_x
u)^2\right],
\label{Hel2d}
\end{equation}
where $K$ and $B$ are tilt and compressional elastic moduli,
respectively, that we take to be phenomenological
parameters.\cite{GLelastic} The parameter $h$ encodes the effect of an
additional magnetic field $H_\perp$, applied transversely to the
columnar defect ($z$-) axis, with $h=H_\perp/(\phi_0 n_0^2)=H_\perp/H_z$
in a sample that (other than the pin) we take for simplicity to be
isotropic.\cite{commentHperp}

Vortex pinning, characterized by a weak potential $V_{\rm pin}(x,z)$
can be easily incorporated through its coupling to the local vortex
density $n_v(x,z)$ via
\begin{equation}
\H_{p}=\int dx dz V_{\rm pin}(x,z) n_v(x,z).
\label{Hp}
\end{equation}
As with any periodic elastic medium, vortex density is given by
\begin{equation}
n_v(x,z)\approx n_{0}-n_0\partial_x u +\sum_{G_p} n_{G_p}\;e^{i G_p(x+u(x,z))},
\label{nv}
\end{equation}
with vortex lattice distortion $u(x,z)$ entering through the variation
of the long-scale density fluctuation, $-n_0\partial_x u$, and via the
variation of the phase, $G_p u$ of the vortex density wave given by
the last term. In above, $G_p=2\pi n_0 p= 2\pi p/a$ ($a$ the vortex
lattice constant, $p\in Z$) spans a one-dimensional reciprocal (to
$x$) lattice and $n_0=1/a$ the average $x$-projected vortex 1D
density. Above representation for $n_v(x,z)$ can be derived in a
standard way from its microscopic definition $n_v(x,z) =
\sum_i\delta(x-x_i(z))$ in terms of vortex-line configurations
$x_i(z)$, by the use of the Poisson summation formula\cite{HaldaneLL},
with the key periodic (last) term arising from vortex discreteness.

For the problem of a single $z$-directed columnar defect, we can
approximate the pinning potential by an attractive zero-range form
$V_{\rm pin}=-V_0 c\delta(x)$, with $V_0$ and $c$ its effective
strength and range, respectively. For simplicity, and without loss of
qualitative generality, we include only the lowest harmonic,
characterized by the minimal reciprocal lattice constants $G_{\pm
1}\equiv\pm G=\pm 2\pi/a$ to model the periodic vortex
density. Furthermore, by minimizing the total energy, it is easy to
show that the long-range part of the density variation,
$-n_0\partial_x u$ has a simple effect of a small shift in vortex
positions, $u_0(x,z) = - {V_0 n_G c \over B}sgn(x)$, that is constant
and positive to the left of the defect and constant and negative to
the right of the defect. It thereby slightly increases the average
vortex density, but only at the location of the columnar defect,
$x=0$, that can be absorbed into the background density.  Dropping an
unimportant constant, the pinning Hamiltonian then reduces to
\begin{equation}
\H_{p}=- v \int dz \cos(G u(0,z)),
\label{Hp2}
\end{equation}
localized at the defect at $x=0$, with $v\equiv 2 n_G V_0 c$. The
resulting total Hamiltonian, $\H=\H_{el}+\H_p$
\begin{eqnarray}
\H&=&{1\over2}\int dx dz\left[K(\partial_z u - h)^2 + B(\partial_x
u)^2\right]\nonumber\\
&&- v \int dz \cos\big(G u(0,z)\big),
\label{H}
\end{eqnarray}
is reminiscent of the well-known sine-Gordon model describing a broad
spectrum of commensurability phenomena in condensed matter physics,
ranging from crystal surface roughness to topological defects in
ordered media\cite{roughening,Pokrovsky,Coppersmith}

There is, however, an essential difference in that the nonlinear
pinning term is localized at $x=0$.  As a result, away from the
defect, the system is harmonic and therefore solvable by elementary
methods. Clearly, as illustrated in Fig.\ref{2DlatticeTilted} away
from the defect the vortex lattice must asymptote to that of a
columnar defect-free configuration, that simply follows the transverse
field
\begin{equation}
u_\infty(x,z)= h z,\ \ \ {\rm for}\ x\rightarrow \pm\infty,
\label{u_infty}
\end{equation}
obtained by minimizing $\H$ for $v=0$. As we will show below, we can
fruitfully take advantage of the locality of the pinning potential by
``integrating out'' (eliminating) the bulk elastic degrees of freedom
away from the defect, thereby reducing the two-dimensional problem to
an effective one-dimensional nonlinear one, that can be solved
exactly. The approach is quite similar to Kane-Fisher's treatment of a
point impurity in a one-dimensional electron liquid.\cite{KaneFisher}

\subsection{Reduction to one-dimensional model}
\label{Reduction1d}

As noted above, because of the short-range nature of the pinning
potential in $\H$, Eq.\ref{H}, vortex degrees of freedom, $u(x,z)$ away
from the columnar defect at $x=0$, are governed by a {\em harmonic}
Hamiltonian. As a result, they are simply related to vortex distortion
at the columnar defect, allowing us to eliminate $u(x,z)$ in favor of
$u(0,z)$.  To automatically satisfy the boundary conditions
$\partial_z u(x,z)|_{z=0,L}=h$, induced by finite $h$, it is
convenient to shift to a new phonon variable, measuring vortex lattice
distortion relative to $u_\infty$,
\begin{equation}
\tilde{u}(x,z)=u(x,z) - h z,
\end{equation}
in terms of which the Hamiltonian becomes
\begin{eqnarray}
\H&=&{1\over2}\int dx dz\left[K(\partial_z\tilde{u})^2 + B(\partial_x
\tilde{u})^2\right]\nonumber\\
&&- v \int dz \cos[G(\tilde{u}(0,z)+h z)].
\label{Htilde}
\end{eqnarray}
At zero temperature a reduction to an effectively one-dimensional
model can be most straightforwardly done by solving the Euler-Lagrange
equation for $u(x,z)$ for a prescribed (but arbitrary) distortion
$u(0,z)\equiv u_0(z)$ on the columnar defect (In Appendix A, we present
a complementary constrained functional integral-based derivation of
above result, that extends to finite temperature)
\begin{equation}
(K\partial_z^2 + B\partial_x^2)u(x,z) = B u_0(z)\partial_x\delta(x).
\label{ELuxz}
\end{equation}
The local stress term (source term on the right-hand-side) is chosen
so as to produce a vortex lattice distortion at $x=0$ to automatically
satisfy the boundary condition $u(0,z)\equiv u_0(z)$. Standard Fourier
analysis leads to a solution
\begin{equation}
\tilde{u}(x,q_z) = \tilde{u}_0(q_z) e^{-(K/B)^{1/2}|q_z||x|},
\label{uxq}
\end{equation}
which, when substituted into the Hamiltonian, $\H$,
Eq.\ref{Htilde}, and integrating over $x$ reduces to
\begin{eqnarray}
\H_{0}&=&\sqrt{K B}\int {d q_z\over 2\pi}|q_z||\tilde{u}_0(q_z)|^2\nonumber\\
&&- v \int dz \cos[G(\tilde{u}_0(z)+h z)].
\label{Htilde0}
\end{eqnarray}
The two-dimensional nature of the underlying vortex lattice is
captured by the {\em nonanalytic} form ($|q_z|$) of the effective
one-dimensional elasticity in $\H_0$. As in other examples of
a low-dimensional system coupled to a bulk system (e.g., a crack or a
crystal surface in a bulk solid\cite{elasticityLR}) it encodes
long-range interactions of one-dimensional deformations mediated
through bulk (away from the columnar defect) degrees of freedom, as
can be easily seen by reexpressing $\H_0$ in terms of $\tilde{u}_0(z)$
\begin{eqnarray}
\H_{0}&=&{\sqrt{K B}\over 2\pi}\int\int dz
dz'\left({\tilde{u}_0(z)-\tilde{u}_0(z')\over z-z'}\right)^2\nonumber\\ 
&&- v \int dz\cos[G(\tilde{u}_0(z)+h z)].
\label{Htilde0z}
\end{eqnarray}

The long-range elasticity qualitatively distinguishes this system from
a one-dimensional sine-Gordon model characterized by short-range and
therefore analytic ($q^2$) long-scale elasticity.  As we will see in
Sec.\ref{finiteT}, the associated enhanced stiffening of elastic
distortions is what allows this one-dimensional system to undergo a
finite temperature roughening phase transition, in contrast to a
one-dimensional sine-Gordon model. For reasons that will become clear
below and by the analogy with the sine-Gordon model, we refer to this
system as the sine-Hilbert model. In the next section we will study
the sine-Hilbert model at zero temperature but finite tilting field
$h$, in order to characterize a vortex lattice response to a magnetic
field $H_\perp$ applied transversely to the columnar defect.

\section{Zero-temperature transverse field response}
\label{Hperp}

\subsection{Bulk tilt response}

For $h=0$ much is known about the sine-Hilbert model, $\H_0$,
Eqs.\ref{Htilde0},\ref{Htilde0z}, as it arises in many different
physical contexts including resistively-shunted Josephson
junctions\cite{RSJ} and Luttinger liquid transport in the presence of
an impurity\cite{KaneFisher}.  In the present context, these findings
relate to a vortex lattice at finite temperature, a study that we will
undertake in Sec.\ref{finiteT}.  Furthermore, using inverse scattering
transform Santini, Ablowitz and Fokas\cite{Ablowitz} have shown that
the classical sine-Hilbert model is integrable and admits soliton
solutions, a finding that we will make use of below.

Less is known about the finite $h$ phenomenology, the study of which
is facilitated by returning to the $u_0(z)=\tilde{u}_0(z) + h z$
displacement field. In terms of $u_0(z)$, the Hamiltonian becomes:
\begin{eqnarray}
\H_{0}&=&{\sqrt{K B}\over 2\pi}\int\int dz dz'\left({u_0(z)-u_0(z')-
h(z-z')\over z-z'}\right)^2\nonumber\\ 
&&- v \int dz\cos[G u_0].
\label{H0z}
\end{eqnarray}

For a finite transverse field, a good starting reference point are two
competing solutions 
\begin{eqnarray}
u_C(z)&=&0,\\ 
u_I(z)&=&h z,
\label{uCuI}
\end{eqnarray}
that minimize $\H_0[u_0(z)]$ in the $h\rightarrow 0$ and
$h\rightarrow\infty$ limits, respectively. By analogy with
commensurate-incommensurate (CI) phase
transitions\cite{Pokrovsky,Coppersmith} we refer to the corresponding
phases as ``commensurate'' and ``incommensurate'', respectively. In
the commensurate state, $u_C(z)$, vortex lattice aligns with the
columnar defect, minimizing the pinning energy, while raising the
diamagnetic energy. In the incommensurate state, $u_I(z)$, instead,
the vortex lattice aligns with the external field, thereby ignoring
the defect and sacrificing its attractive pinning energy. For a sample
of extent $L$ along $z$, the corresponding total energies in the two
cases are given by
\begin{mathletters}
\begin{eqnarray}
E_C&=&{\sqrt{BK}\over 2\pi} h^2 L^2 - v L,\label{EC}\\ 
E_I&=&
0.\label{EI}
\end{eqnarray}
\end{mathletters}
The extensive scaling, $\sim L^2$ of the magnetic energy in $E_c$ is
expected from Eq.\ref{uxq}, that shows that elastic distortion on
scale $L$ along $z$, imposed at $x=0$ (in the commensurate phase
corresponding to misalignment with applied field) decays over length
$L (B/K)^{1/2}$ into the bulk, leading to a region of area $\sim
L^2(B/K)^{1/2}$ with a finite diamagnetic energy cost. As discussed in
the Introduction, in the thermodynamic limit this bulk diamagnetic
energy always dominates over the linear-with $L$ pinning energy, and
vortex system is in the incommensurate state for arbitrarily small
transverse field $h$.  However, for a finite $L$, we find from
Eqs.\ref{EC}, \ref{EI}, that for a single pin, the tilting transition
between $u_C$ and $u_I$ states takes place at the critical
(``thermodynamic bulk'') tilting field
\begin{eqnarray}
h_{c}&=&\left({2\pi v\over\sqrt{BK}}\right)^{1/2}{1\over L^{1/2}},
\end{eqnarray}
that, as expected, vanishes in the thermodynamic limit,
$L\rightarrow\infty$. For a finite density of columnar defects spaced
by $d$, it is clear that the bulk critical field saturates at a finite
$d$-dependent value given by
\begin{mathletters}
\begin{eqnarray}
h_c&\approx&
\cases{\left({2\pi v\over\sqrt{BK}}\right)^{1/2}{1\over L^{1/2}},
& $L\sqrt{B\over K}\ll d$\cr 
\left({2\pi v\over K}\right)^{1/2}{1\over d^{1/2}},
& $L\sqrt{B\over K}\gg d$\cr},
\label{hc_limits}
\end{eqnarray}
\end{mathletters}
This oversimplified picture of the tilting transition, ignoring
solitons (see below) is summarized in Fig.\ref{E_ICnaive}.

\begin{figure}[bth]
\centering
\setlength{\unitlength}{1mm}
\begin{picture}(40,70)(0,0)
\put(-22,-2){\begin{picture}(30,40)(0,0)
\includegraphics{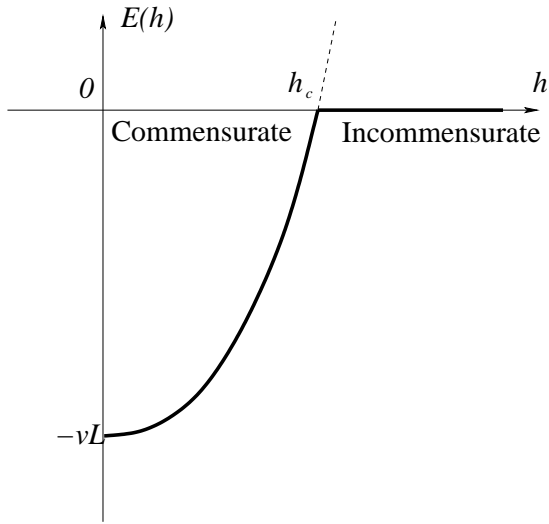}
\end{picture}}
\end{picture}
\caption{Oversimplified (ignoring solitons) 
expression for the energy $E(h)$ of a vortex lattice in the presence
of a transverse field $h$, indicating the aligned (C) to tilted (I)
transition at $h_c$.}
\label{E_ICnaive}
\end{figure}

\subsection{Tilt-solitons}

\begin{figure}[bth]
\centering
\setlength{\unitlength}{1mm}
\begin{picture}(40,75)(0,0)
\put(8,-2){\begin{picture}(30,60)(0,0)
\includegraphics{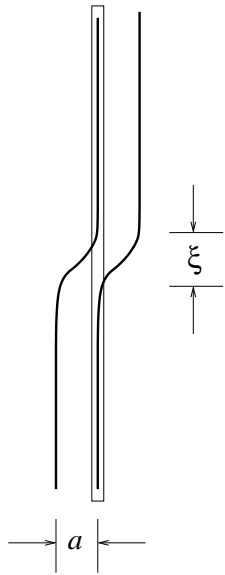}
\end{picture}}
\end{picture}
\caption{A two-dimensional perspective of a 1D soliton along $z$,
corresponding of an exchange between two neighboring vortex lines
localized on the columnar defect.}
\label{soliton2D}
\end{figure}

Two states, $u_{C}$, $u_I$, are only appropriate in the $h\rightarrow
0$, $h\rightarrow\infty$ limits, respectively. However, in analogy
with other systems, where there is competition between elastic and
pinning energies\cite{roughening,Pokrovsky,Coppersmith}, we expect
and find (see below), that the tilting transition into an
incommensurate state is driven by soliton proliferation transition
above a lower-critical field $h_{c1}$, that preempts the bulk
transition at $h_c$ found above.\cite{commentHc} The soliton state
above $h_{c1}$ then continuously approaches the fully incommensurate
$u_I(z)$ solution in the $h\rightarrow\infty$ limit, when solitons
become dense.\cite{commentHc} The existence of a lower-energy soliton
solution can be seen by a simple inspection of the Hamiltonian $\H_0$,
Eq.\ref{H0z}. It stems from the periodicity of the pinning
energy, $\H_{pin}[u_0]=\H_{pin}[u_0+a]$, that microscopically
corresponds to its independence of which of the identical vortex lines
in the array is pinned by the columnar defect. A soliton at $z_0$
corresponds to a solution $u_{0s}(z)$ that switches at $z_0$ between
two adjacent minima of the periodic potential. As illustrated in
Fig.\ref{soliton2D}, from the 2D perspective such soliton describes a
switching between two adjacent vortex lines localized on the columnar
defect.

At zero temperature the soliton shape is characterized by a single
length scale,
\begin{equation}
\xi_0=\left({a\over2\pi}\right)^2{2\sqrt{B K}\over v},
\label{xi0}
\end{equation}
that can be read off from $\H_0$ by balancing the elastic and pinning
energies, that scale like $\sqrt{KB}a^2$ and $v\xi_0$, respectively.
This soliton width is set by the balance between the pinning and
elastic energies. The former (latter) is minimized by the most (least)
abrupt shift between the adjacent pinning minima, with $\xi_0$
reflecting this through its dependence on elastic moduli and pinning
strength, increasing with $B,K$ and decreasing with $v$. We note that
the effective 1D modulus is a geometrical mean of the vortex
compressional (B) and tilt (K) moduli, as 2D distortion corresponding
to the soliton configuration along the defect in Fig.\ref{soliton2D}
involves both tilt and compression of the 2D vortex lattice.

It is convenient to express length scales along $z$ in units of the
soliton width, $\xi_0$, and trade in the displacement field $u_0(z)$
for a dimensionless phase field
\begin{equation}
\phi(\z)={2\pi\over a} u_0(z),
\label{phi}
\end{equation}
where throughout the paper we will used ``hat'ed'' symbols to denote
dimensionless quantities.  The Hamiltonian $\H_0$ then reduces to a
dimensionless form
\begin{eqnarray}
{\H_{0}\over\epsilon_0}&=&{1\over 4\pi}\int\int d\z
d\z'\left({\phi(\z)-\phi(\z') - \hat{h}(\z-\z')\over \z-\z'}\right)^2
\nonumber\\ 
&&-\int d\z\cos[\phi],
\label{Htilde0dim}
\end{eqnarray}
where 
\begin{eqnarray}
\epsilon_0&=&\left({a\over2\pi}\right)^2 2\sqrt{B K},\\
&=& \xi_0 v,
\end{eqnarray}
is the soliton energy scale and 
\begin{equation}
\h\equiv{2\pi\xi_0\over a}h
\end{equation}
is a dimensionless measure of a transverse field.

These energy and length scales are sufficient to determine all
qualitative ingredients of the soliton-driven tilting (CI) transition.
However, integrability of the Euler-Lagrange equation ${\delta
\H_0\over\delta u_0(z)}=0$
\begin{eqnarray}
{1\over\pi}\int d\z' {\phi(\z) - \phi(\z')\over(\z-\z')^2} + \sin\phi(\z)=0,
\label{eom}
\end{eqnarray}
found by Santini, et al.\cite{Ablowitz} allows for a quantitatively
exact analysis. In above, the integral must be defined as the
principle value with the singular point $z=z'$ excluded, and we have
used free boundary conditions on $\phi(L)$ and $\phi(0)$ in deriving
Eq.\ref{eom}. These lead to a boundary condition
\begin{equation}
\partial_{\z}\phi(\z)|_{\z=0,\L}=\h,
\end{equation}
that supplements Eq.\ref{eom}. It encodes the condition that at the
edge of the sample vortex lines tilt to slope $h$ to follow the
external magnetic field.

\begin{figure}[bth]
\centering
\setlength{\unitlength}{1mm}
\begin{picture}(40,60)(0,0)
\put(-25,-2){\begin{picture}(30,50)(0,0)
\includegraphics{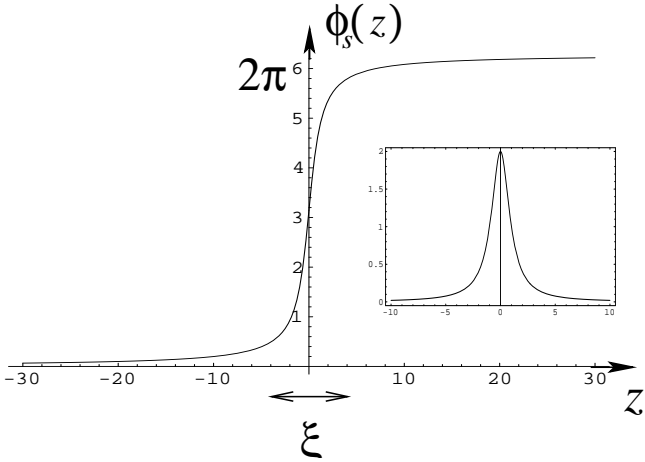}
\end{picture}}
%\put(-20,40) {$n_p$}
%\put(60,2) {$\epsilon_p$}
\end{picture}
\caption{Soliton solution $\phi_s(z)$, Eq.\ref{soliton} to the sine-Hilbert 
equation, Eq.\ref{eom}. Inset shows $\partial_z\phi_s(z)$.}
\label{solitonFig}
\end{figure}

A single-soliton solution to the above sine-Hilbert equation,
Eq.\ref{eom} was discovered by Rudolf Peierls\cite{Peierls} in his
seminal study of an edge dislocation in a crystal, and later
rediscovered and considerably extended (to multi-solitons, dynamics
and proving integrability) using the inverse scattering method by
Santini, et al.\cite{Ablowitz}. It is illustrated in
Fig.\ref{solitonFig} and given by
\begin{equation}
\phi_s(\z)=-2{\rm ArcTan}{1\over \z-\z_0},
\label{soliton}
\end{equation}
This solution can be verified by a direct substitution of $\phi_s(z)$
into Eq.\ref{eom}, using Hilbert transforms that we review in Appendix
\ref{singleSoliton}.

Substituting $\phi_s(\z)$ into $\H_0$, Eq.\ref{Htilde0dim}, and using
Hilbert transforms (with details presented in Appendix
\ref{solitonEnergy}), we find the energy of a single soliton state
\begin{eqnarray}
E_{1}&=&\H_{0}[\phi_s(\z)],\\ &\equiv&\epsilon_0\E_C+\epsilon_0\E_{s1}
\label{Es1}
\end{eqnarray}
where dimensionless single soliton energy, $\E_{s1}$, computed in
Appendix \ref{solitonEnergy} is given by
\begin{eqnarray}
\E_{s1}&=&2\pi\ln\big({e\L\over 4}\big)-\h(2\L-2\pi),
\label{Es1dim}
\end{eqnarray}
$\L\equiv L/\xi_0$, and $\E_C$ the dimensionless energy of the
commensurate state (cf. Eq.\ref{EC})
\begin{equation}
\E_C={1\over4\pi}\h^2\L^2- \L.
\label{hatEC}
\end{equation}

From $\E_{s1}$, Eq.\ref{Es1dim} it is clear that soliton energy
becomes negative for $\h > \h_{c1}$, with
\begin{eqnarray}
h_{c1}&=&{a\over\xi_0}{1\over 2\L-2\pi}\ln{e \L\over2},\\
&\approx&{a\over 2L}\ln{L\over\xi_0},
\label{hc1}
\end{eqnarray}
that, as expected (given that there is only a single pin) vanishes in
the thermodynamic limit.  As with other CI phase
transitions\cite{Pokrovsky,Coppersmith}, this leads to soliton
proliferation at $h_{c1}$\cite{commentHc}, that preempts the bulk
transition at $h_c$, approaching the fully tilted incommensurate state
$u_I(z)$, only as $h\rightarrow h_{c2}$. The latter tilt field is
defined by when the solitons begin to overlap, namely
$n_s(h_{c2})\approx\xi_0^{-1}$. However, in contrast to conventional
CI transition, where the ratio of the lower critical field to the
thermodynamic one is of order $1$ constant ($=4/(\sqrt{2}\pi)\approx
0.90$ for the CI transition in the sine-Gordon model), here
\begin{equation}
{h_{c1}\over h_{c}}={\pi^{1/2}\over2}\left({\xi_0\over
L}\right)^{1/2}\ln{L\over\xi_0}.
\end{equation}
Vanishing of this ratio for large $L$ demonstrates the importance of
solitons in driving the vortex lattice tilting transitions.

\subsection{Tilt-soliton proliferation transition}

\begin{figure}[bth]
\centering
\setlength{\unitlength}{1mm}
\begin{picture}(40,60)(0,0)
\put(-25,-2){\begin{picture}(30,40)(0,0)
\includegraphics{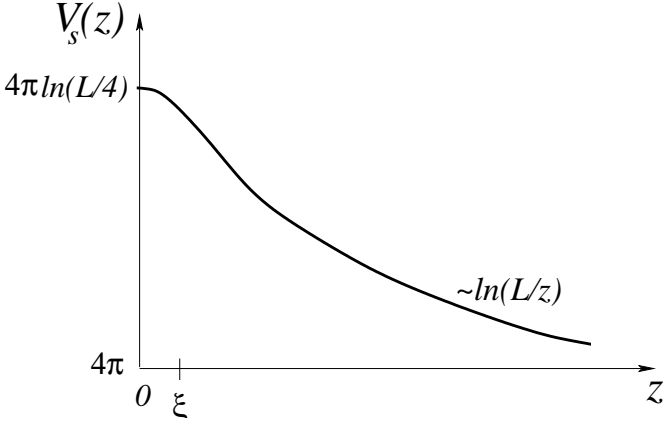}
\end{picture}}
%\put(-20,40) {$n_p$}
%\put(60,2) {$\epsilon_p$}
\end{picture}
\caption{Soliton interaction potential $V_s(z)$, 
illustrating a long-range logarithmic repulsion.}
\label{VsFig}
\end{figure}

The vortex lattice tilting angle, $\theta(h)$, related to transverse
flux density $B_\perp(H_\perp)$, via $b_\perp\equiv\tan\theta =
B_\perp/B_z$ is determined by the soliton density $n_s(h)$ through the
relation
\begin{equation}
b_\perp(h)= a n_s(h)
\end{equation}
The soliton density $n_s(h)$ for $h>h_{c1}$ in turn is determined by
the balance of the soliton chemical potential energy $h$ (that induces
solitons) and the soliton repulsive interaction.  In the dilute
soliton limit (for $h_{c1} <h \ll h_{c2}$), we can approximate the
latter by a sum of pair-wise soliton interactions. This is determined
by the energy $E_{2}(\z_1,\z_2)\equiv\epsilon_0\E_{2}(\z_1,\z_2)$ of
two one-solitons, localized at $\z_1$ and $\z_2$, separated by a large
distance $\z_1-\z_2\gg 1$, with
\begin{widetext}
\begin{eqnarray}
\hspace{-1cm}\E_{2}(\z_1,\z_2)&\equiv&\Hh_0[\phi_{s1}+\phi_{s2}],\\
&=&{1\over 4\pi}\int\int d\z
d\z'\left({\phi_{s1}(\z)+\phi_{s2}(\z)-\phi_{s1}(\z')-\phi_{s2}(\z')-
\h(\z-\z')\over \z-\z'}\right)^2-\int d\z\cos[\phi_{s1}(\z)+\phi_{s2}(\z)].
\label{E2sDefine}
\end{eqnarray}
\end{widetext}
A straightforward calculation along the lines of the computation of
$E_{1}$ gives (with details presented in Appendix
\ref{soliton2Energy}):
\begin{eqnarray}
\hspace{-0.5cm}
\E_{2}(\z_1,\z_2)&=&\E_C+2\E_{s1}(\h)+V_s(\z_1-\z_2),
\end{eqnarray}
where the soliton two-body repulsive interaction is given by
\begin{eqnarray}
\hspace{-0.5cm}
V_s(\z)&=&2\pi\ln\left[{(\L/2)^2\over \z^2+4}\right] + 4\pi,\\
&\approx& 4\pi\ln{\L\over 2|\z|}+4\pi,\; {\rm for}\; 1\ll |\z|\ll \L/2,
\label{V_s}
\end{eqnarray}
and is illustrated in Fig.\ref{VsFig}. 

The energy of the $N_s$ soliton lattice is then given by
\begin{eqnarray}
\hspace{-1.5cm}
\E_{Ns}&=&\E_C + N_s \E_{s1} + \sum_{i<j} V_s(\z_i-\z_j),\\
&\approx& \E_C + 2\L (\h_{c1}-\h) N_s +{1\over2} V_s(\L/2)N_s^2,\\ 
&\approx&\E_C + \L^2\left[2(\h_{c1}-\h)\n_s +{1\over2} V_s(\L/2)\n_s^2\right],
\qquad\label{ENs}
\end{eqnarray}
where, because of the long-range nature (logarithmic) of the
interaction $V_s(\z)$, the sum in the interaction energy is dominated
by the longest length scale, $L$, leading to its $N_s^2=n_s^2 L^2$
stronger-than-extensive growth. This is in qualitative contrast to
standard CI $T=0$ transitions\cite{Pokrovsky,Coppersmith,commentHc},
where the interaction is short-ranged and the corresponding sum is
dominated by the smallest term $V_s(a/\xi_0)$, leading to an extensive
interaction energy.\cite{logInteraction}

Minimizing $\E_{Ns}[\n_s]$ over the soliton density $\n_s=N_s/\L$, we
find the advertised CI soliton proliferation
transition\cite{noTransition} at $\h_{c1}$, with the soliton density
\begin{eqnarray}
\n_s(\h) = \cases{0,& $\h < \h_{c1}$\cr 
\n_{s0}(\h-\h_{c1}), & $\h > \h_{c1}$,\cr}
\label{ns}
\end{eqnarray}
growing linearly with the transverse field $h$. $\n_{s0}$ is $O(1)$
constant given by
\begin{eqnarray}
\n_{s0}&\equiv& {2\over V_s(\L/2)},\\
&\approx& {1\over 2\pi},\label{estimate}
\end{eqnarray}
The final numerical result, Eq.\ref{estimate}, above is quite crude,
providing only an order of magnitude estimate, as we have not
carefully treated the case of a soliton near the edge of the system.

\begin{figure}[bth]
\centering
\setlength{\unitlength}{1mm}
\begin{picture}(40,85)(0,0)
\put(-20,3){\begin{picture}(30,80)(0,0)
\includegraphics{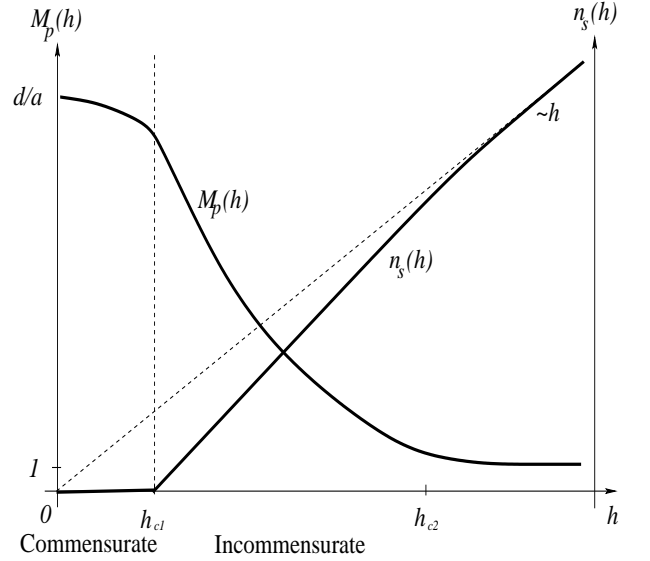}
\end{picture}}
\end{picture}
\caption{A sketch of soliton density, $n_s(h)$, and the number of 
vortex lines pinned per columnar defect, $M_p(h)$, as functions of the
transverse field $h$.}
\label{ns_hFig}
\end{figure}

The dimensionless soliton density $\n_s$ (measured in units of
$1/\xi_0$) increases to $O(1)$ when $\h=h {2\pi\xi_0\over a}$ exceeds
$\h_{c1}$ by $1$. This corresponds to the physical soliton density
$n_s(h_{c2})\approx 1/\xi_0$ of a dense lattice of overlapping
solitons, with the upper-critical transverse field $h_{c2}$ given by
\begin{eqnarray}
h_{c2}=h_{c1}+ {a\over 2\pi\xi_0},
\label{hc2}
\end{eqnarray}
consistent with physical picture illustrated in Fig.\ref{solitonNFig}
and summarized by Fig.\ref{ns_hFig}

\begin{figure}[bth]
\centering
\setlength{\unitlength}{1mm}
\begin{picture}(40,70)(0,0)
\put(-22,-2){\begin{picture}(30,40)(0,0)
\includegraphics{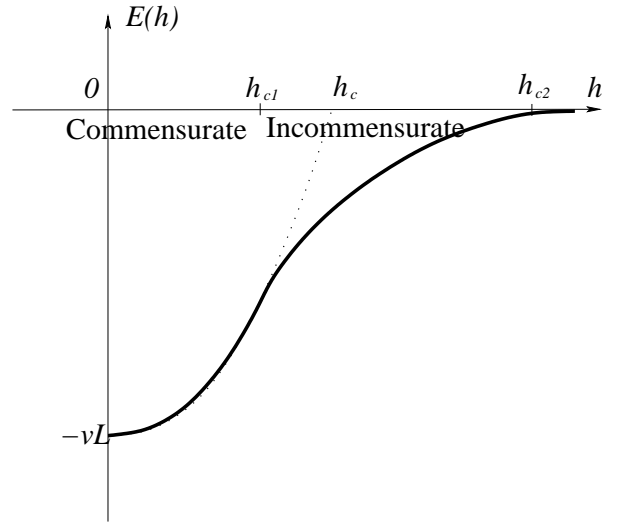}
\end{picture}}
\end{picture}
\caption{Vortex lattice energy $E(h)$ as a function of transverse field $h$, 
illustrating soliton-driven tilting CI transition at the
lower-critical field $h_{c1}$ preempting the bulk thermodynamic
critical field $h_c$. At high upper-critical transverse field $h_{c2}$
the system crosses over to a fully discommensurated tilted state with a
smooth tilt response.}
\label{E_ICfig}
\end{figure}

Using solution $N_s(\h)= \n_s(\h)\L$, Eq.\ref{ns} inside $\E_{Ns}$ we
find the vortex lattice energy as a function of the transverse field
$h$:
\begin{eqnarray}
\hspace{-1.2cm}\E(\h)= \cases{{\h^2\L^2\over4\pi}- \L,& $\h < \h_{c1}$\cr 
{\h^2\L^2\over4\pi} - \L - {\n_{s0}\over 2}(\h-\h_{c1})^2\L^2, &
$\h_{c1}<\h\ll \h_{c2}$.\cr}\nonumber
\hspace{-2cm}&&\\
\label{E_Ns}
\end{eqnarray}
As expected the negative soliton energy (last term) cancels off the
$\L^2$ misalignment-field energy (first term), as the incommensurate
state is approached with increasing soliton density in the large
$\h\gg \h_{c1}$ limit.\cite{commentCancel} This energy and the
corresponding phase diagram are illustrated in Fig.\ref{E_ICfig}.

\subsection{Bulk vortex lattice distortions}
\label{bulk_distortion}

\begin{figure}[bth]
\centering
\setlength{\unitlength}{1mm}
\begin{picture}(40,40)(0,0)
\put(-22,0){\begin{picture}(30,40)(0,0)
\includegraphics{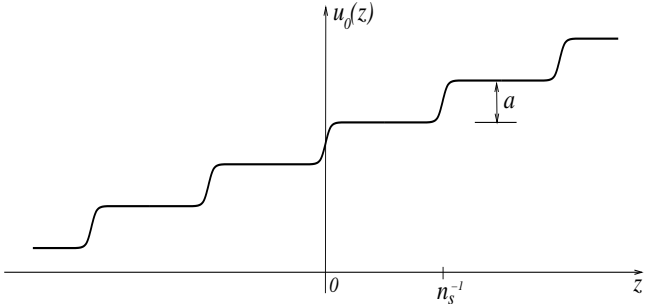}
\end{picture}}
\end{picture}
\caption{A sketch of the function $u_0(z)$ for an array of 5 solitons.}
\label{u0Fig}
\end{figure}

Having established the existence and computed the details of the $T=0$
vortex array tilting transition in terms of the one-dimensional field
$u_0(z)$, illustrated in Fig.\ref{u0Fig}, we now turn to the
computation of the associated bulk vortex lattice distortion
$u_0(x,z)$.
The connection between $u_0(z)$ and $u_0(x,z)$ is established through
a key relation, Eq.\ref{uxq} (here we use the subscript $0$ to denote
$T=0$ field configurations, obtained by minimizing the energy)
\begin{equation}
\tilde{u}_0(x,q_z) = \tilde{u}_0(q_z) e^{-|x|/\lambda_h(q_z)},
\label{uxq2}
\end{equation}
that shows quite clearly that a one-dimensional distortion with a
wavevector $q_z$ along the defect penetrates into the bulk $x\neq 0$
over a screening length
\begin{equation}
\lambda_h(q_z)=\left({B\over K}\right)^{1/2}{1\over |q_z|}.
\label{lambda}
\end{equation}

\begin{figure}[bth]
\centering
\setlength{\unitlength}{1mm}
\begin{picture}(40,15)(0,0)
\put(-23,3){\begin{picture}(30,15)(0,0)
\includegraphics{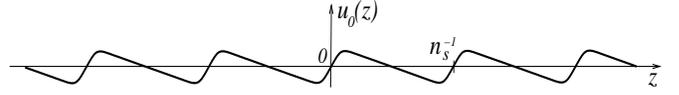}
\end{picture}}
\end{picture}
\caption{A sketch of the function $\overline{u}_0(z)$, Eq.\ref{u0sbar}, 
characterizing the periodic part of the soliton train solution
$u_0$.}
\label{u0barFig}
\end{figure}

A real-space one-dimensional distortion in an $N_s$ soliton state is
given by
\begin{eqnarray}
\tilde{u}_0(z)&=&\sum_{\alpha=1}^{N_s} u_s(z-z_\alpha)-h z,\\
&\equiv&\overline{u}_0(z) - (h-n_s a) z,
\label{u0s}
\end{eqnarray}
with $u_s(z)=(a/2\pi)\phi_s(z)$ the single soliton solution from
Eq.\ref{soliton}.  Physically $\partial_z\tilde{u}_0(z)$ is
proportional to transverse magnetization. In Eq.\ref{u0s} we defined a
periodic part coming from the soliton array (illustrated in
Fig.\ref{u0barFig})
\begin{eqnarray}
\overline{u}_0(z)&=&\sum_{\alpha=1}^{N_s} u_s(z-z_\alpha)- n_s a z,
\label{u0sbar}
\end{eqnarray}
that oscillates around $0$ with period $1/n_s$,
and the remaining misalignment part $h_{\rm eff} z$, with an effective
tilt-field
\begin{eqnarray}
h_{\rm eff}&=&h-n_s(h)a,\\
&=& h - 2\pi\n_{s0}|h-h_{c1}|.
\label{heff}
\end{eqnarray}
These respectively contribute to the oscillatory and uniform part of
the (negative) transverse magnetization. 

\begin{figure}[bth]
\centering
\setlength{\unitlength}{1mm}
\begin{picture}(40,25)(0,0)
\put(-23,3){\begin{picture}(30,25)(0,0)
\includegraphics{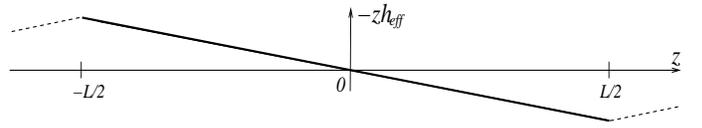}
\end{picture}}
\end{picture}
\caption{A sketch of a tilted component of $u_0(z)$, periodically extended 
beyond the system size $L$.}
\label{heffzFig}
\end{figure}

To compute the Fourier transform $\tilde{u}_0(q_z)$, we extended
$h_{\rm eff} z$ part of $\tilde{u}_0(z)$ beyond the system size $L$ to
a continuous periodic function with period $2L$ illustrated in
Fig.\ref{heffzFig}.

From this decomposition it is clear that the distortion
$\tilde{u}_0(z)$ appears on two characteristic wavelengths: the system
size $L$ along $z$ (we take the system to be infinite along $x$) and
the soliton spacing $n_s^{-1}$.  Hence we predict that the associated
bulk distortion, $u_0(x,z)$ will display an average misalignment with
the applied field (i.e., on average align with the columnar defect),
that extends over a length
\begin{equation}
\lambda_h^{0}=\left({B\over K}\right)^{1/2}{L\over\pi},
\label{lambda_tilt}
\end{equation}
along $x$ and will exhibit soliton-induced oscillations with
wavelength $\sim n_s^{-1}$ that penetrate over scale
\begin{equation}
\lambda_h=\left({B\over K}\right)^{1/2}{1\over2\pi n_s(h)},
\label{lambda_h}
\end{equation}
into the bulk, $x\neq 0$ away from the columnar defect.

Above qualitative discussion can be elevated to an exact
calculation. Using Eq.\ref{uxq2}, the real-space bulk distortion is
given by
\begin{equation} 
u_0(x,z) = h z + \int{d q_z\over 2\pi} \tilde{u}_0(q_z)
e^{-({K\over B})^{1\over2}|q_z||x|} e^{i q_z z}.
\label{uxz}
\end{equation}
Putting together above ingredients, the Fourier transform
$\tilde{u}_0(q_z)$ can be easily computed, and, when inserted into
Eq.\ref{uxz} gives
\begin{widetext}
\begin{eqnarray}
&&\label{uxz2a}\\
u_0(x,z)&=& h z - {4\over\pi^2} L h_{\rm eff}
\sum_{m=0}^\infty{(-1)^m\over(2m+1)^2}
e^{-\sqrt{K\over B}{\pi\over L}(2m+1)|x|}\sin[{\pi\over L}(2m+1)z]
+{a\over\pi}\sum_{p=1}^\infty{1\over p}
e^{-\sqrt{K\over B} 2\pi n_s p|x|}\sin[2\pi n_s p z],\nonumber\\
&\approx& 
\left[1-(1-{n_s(h) a\over h})e^{-\sqrt{K\over B}{\pi\over L}|x|}\right] h z\;
+{a\over\pi}\;e^{-\sqrt{K\over B} 2\pi n_s|x|}\sin[2\pi n_s z].
\label{uxz2b}
\end{eqnarray}
\end{widetext}

\begin{figure}[bth]
\centering
\setlength{\unitlength}{1mm}
\begin{picture}(40,50)(0,0)
\put(-15,2){\begin{picture}(30,40)(0,0)
\includegraphics{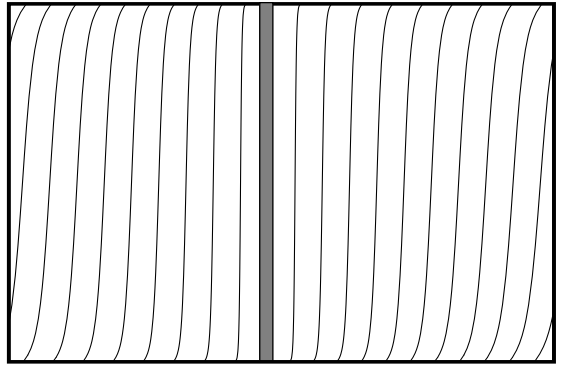}
\end{picture}}
%\put(-20,40) {$n_p$}
%\put(60,2) {$\epsilon_p$}
\end{picture}
\caption{Vortex lattice ($T=0$) configuration $u_0(x,z)$ for a subcritical 
tilt field $h<h_{c1}$, showing expulsion of tilt over a region of size
$L\times (B/K)^{1/2} L$ around the pin, but with penetration of tilt
on scale $\xi_0$ at $x=0$ growing with $|x|$ to full penetration
beyond $|x|\approx (B/K)^{1/2}L$.}
\label{soliton0Fig}
\end{figure}

\begin{figure}[bth]
\centering
\setlength{\unitlength}{1mm}
\begin{picture}(40,50)(0,0)
\put(-15,2){\begin{picture}(30,40)(0,0)
\includegraphics{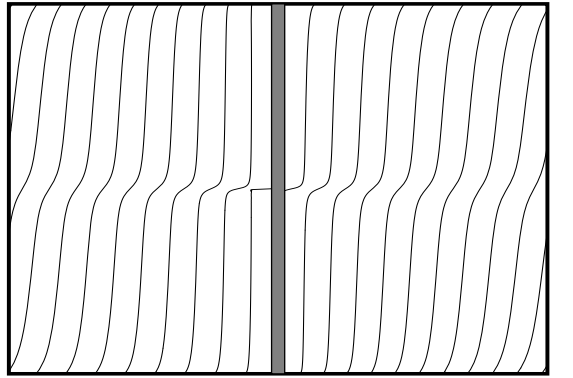}
\end{picture}}
%\put(-20,40) {$n_p$}
%\put(60,2) {$\epsilon_p$}
\end{picture}
\caption{Vortex lattice ($T=0$) configuration $u_0(x,z)$ for just-above 
critical tilt-field $h = h_{c1}^+$, with a single soliton along the
pin. It shows the expulsion of tilt over a region $L\times (B/K)^{1/2}
L/2$ around the pin, with the tilt penetration length
$\lambda_{h2}=(B/K)^{1/2} L/2$ along $|x|$ reduced by a factor of $2$
relative to the case of no solitons, illustrated in
Fig.\ref{soliton0Fig}.}
\label{soliton1Fig}
\end{figure}

\begin{figure}[bth]
\centering
\setlength{\unitlength}{1mm}
\begin{picture}(40,50)(0,0)
\put(-15,2){\begin{picture}(30,40)(0,0)
\includegraphics{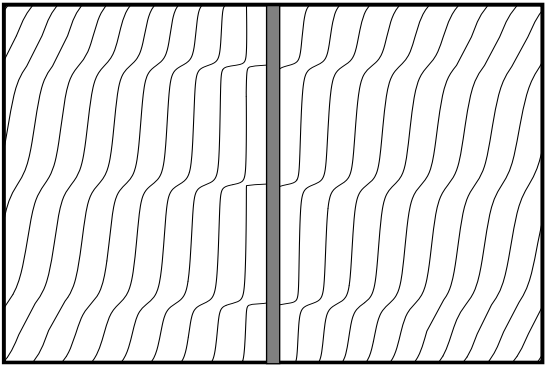}
\end{picture}}
%\put(-20,40) {$n_p$}
%\put(60,2) {$\epsilon_p$}
\end{picture}
\caption{Vortex lattice ($T=0$) configuration $u_0(x,z)$ for tilt-field 
$h_{c1} < h \ll h_{c2}$, with three solitons along the pin. It shows
the expulsion of tilt over a region $L\times (B/K)^{1/2} L/4$ around
the pin, with the tilt penetration length $\lambda_{h4}=(B/K)^{1/2}
L/4$ along $|x|$ reduced by a factor of $4$ relative to the case of no
solitons, illustrated in Fig.\ref{soliton0Fig}. As discussed in the
text, in general, the $N$-soliton penetration length is given by
$\lambda_{h N}=({B\over K})^{1/2} L/N_s\approx
\lambda_{h}/N_s=n_s^{-1}(B/K)^{1/2}$.}
\label{soliton3Fig}
\end{figure}

\begin{figure}[bth]
\centering
\setlength{\unitlength}{1mm}
\begin{picture}(40,50)(0,0)
\put(-15,2){\begin{picture}(30,40)(0,0)
\includegraphics{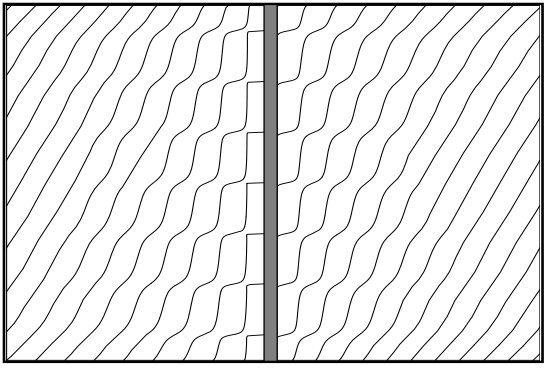}
\end{picture}}
%\put(-20,40) {$n_p$}
%\put(60,2) {$\epsilon_p$}
\end{picture}
\caption{Vortex lattice ($T=0$) configuration $u_0(x,z)$ for tilt-field 
$h\rightarrow h_{c2}$, giving a dense soliton array and a vanishing
tilt-expulsion length. A resulting vortex lattice tilt response is
nearly that of a pin-free system, with the distortion confined to the
immediate vicinity of the columnar defect.}
\label{solitonNFig}
\end{figure}

Although no approximation is required to compute $u_0(x,z)$,
Eq.\ref{uxz}, in Eq.\ref{uxz2a} we have considerably simplified the
Fourier transform by approximating the soliton part
$\overline{u}_0(z)$ by a function $-n_s a z$, periodically extended
with period $n_s^{-1}$. This approximation is valid away from the
dense soliton limit, i.e., for $h\ll h_{c2}$, where $n_s a\ll 1$.  In
going to Eq.\ref{uxz2b} we have furthermore simplified $u_0(x,z)$ by
keeping only the first harmonic in each of the two terms and
approximated the first sine by a line with a proper slope (determined
by boundary conditions) valid for $z\ll L$ away from the edges of the
sample.  The corresponding configurations of the vortex lattice in the
commensurate state, $h<h_{c1}$, slightly inside the soliton state,
$h\approx h_{c1}^{+}$, intermediate regime $h_{c1}<h\ll h_{c2}$, and
deep in the soliton state $h\gg h_{c1}$ are illustrated in
Figs.\ref{soliton0Fig}, \ref{soliton1Fig},
\ref{soliton3Fig}, and \ref{solitonNFig} respectively.

As anticipated by the qualitative discussion above, influence of the
columnar defect depends on the tilt field $h$ and ranges from the
maximum value $\lambda_h^{0}$, Eq.\ref{lambda_tilt} in the
commensurate state, down to $\lambda_h$, Eq.\ref{lambda_h} in the
soliton state. Upon increasing $h$ beyond the lower-critical field
$h_{c1}$, the soliton density increases and the penetration length
decreases according to
\begin{eqnarray}
\hspace{-0.5cm}
\lambda_h(h)&=&{a\over 4\pi^2 \n_{s0}}\left({B\over K}\right)^{1/2}
{1\over |h-h_{c1}|},\label{lambda_h2}\\ 
&\rightarrow&\lambda_h^\infty\equiv {{a\over2\pi}
\left({B\over K}\right)^{1/2}}{1\over h},
\ \ {\rm for}\ \ h_{c1}\ll h\ll h_{c2},\nonumber\\
\label{lambda_hinfty}
\end{eqnarray}
saturating at the microscopic lattice scale $a$ as $h_{c2}$ is
approached. The size of the pinned vortex ``cloud'', $\lambda_h$
allows to define an important dimensionless number
$M_p(h)$\cite{Affleck,commentNp}
\begin{eqnarray}
M_p(h)={\lambda_h\over a},
\end{eqnarray}
that gives the number of vortex lines effectively pinned by a single
columnar defect. As illustrated in Fig.\ref{ns_hFig} it ranges from
$M_p^{\rm max}\approx(B/K)^{1/2}L/a$ ($d/a$ for finite density of
pins) in the commensurate ($h<h_{c1}$) state down to its minimum value
of $1$ for $h\rightarrow h_{c2}$.

\section{Finite temperature thermodynamics and correlations}
\label{finiteT}

Having established the vortex lattice $T=0$ tilt-response to a
transverse magnetic field, we now turn to the study of
finite-temperature properties of the vortex lattice pinned by a single
columnar defect.\cite{Affleck} As usual, the thermodynamics and
corresponding correlation functions can be computed by integrating the
vortex phonon configurations weighted by a Boltzmann factor,
$e^{-\H/{k_B T}}/Z$, with $Z$ the corresponding partition function
$Z=\int [du] e^{-\H/{k_B T}}$.

\subsection{Thermal depinning transition for vanishing tilt-field, $h=0$}
\label{depinningTransition}

We first consider the case of a vanishing tilt-field, $h=0$, and show
that the effective one-dimensional, long-range interacting
Hamiltonian, $\H_0$, to which our 2D vortex problem has been reduced,
exhibits a thermal depinning transition, qualitatively similar to a
well-known roughening transition in 2D
systems.\cite{roughening,Pokrovsky,Coppersmith}

The indication of the existence of the transition comes from a
computation of the thermodynamics perturbatively in powers of the
pinning potential strength $v$. As for the 2D roughening (and related)
transition, this perturbation theory diverges (at long length scales)
for low temperatures $T<T_p$ even for an arbitrarily small $v$, but is
convergent for high temperatures $T>T_p$, with the critical pinning
temperature $T_p$ defined below. As usual to make sense of the
associated infrared divergences near and below $T_p$ we employ the
momentum-shell renormalization-group (RG)
transformation\cite{WilsonKogut} on the 1D Hamiltonian $\H_0$
\begin{eqnarray}
\H_{0}&=&\sqrt{K B}\int {d q_z\over 2\pi}|q_z||\tilde{u}_0(q_z)|^2
- v \int dz \cos[G\tilde{u}_0(z)].\nonumber\\
&&
\label{Htilde0b}
\end{eqnarray}
Namely, in the partition function for the model, we integrate out
perturbatively in $v$ short-scale phonon modes $u_0^>(q_z)$ in a
narrow shell $G_z e^{-\delta\ell}<|q_z|< G_z$ around the short scale
cutoff $G_z=G\sqrt{B/K}\equiv (2\pi/a_z)$, with $\delta\ell\ll
1$. This allows us to express the partition function in terms of
integrals over the remaining effective course-grained phonon modes
$u_0^<(q_z)$, with $|q_z|< G_z e^{-\delta\ell}$. In order to massage
the resulting Hamiltonian into the original form $\H_0$, i.e., to bring
the reduced ultra-violet (UV) cutoff back up to $G_z$, we rescale
lengths and wavevectors according to
\begin{eqnarray}
q_z &=&  q_z' e^{-\delta\ell},\\
z &=& z' e^{\delta\ell}.
\end{eqnarray}
For convenience, to keep the vortex lattice reciprocal vector $G$ in
the argument of the pinning potential fixed, we choose not to rescale
the real-space phonon field $u_0^<(z) = u_0'(z')$, which implies a
rescaling
\begin{eqnarray}
u_0^<(q) = e^{-\delta\ell}u_0'(q'),
\end{eqnarray}
for its Fourier transform. With these transformations the effective
coarse-grained Hamiltonian returns to its original $\H_0$ form, but
with effective, $\ell$-dependent elastic and pinning parameters. As
can be easily checked, to lowest order in $v$, the one-dimensional
stiffness $\sqrt{B K}$ remains unchanged,\cite{commentRGno} and the
effective pinning strength transforms according to
\begin{eqnarray}
v(\delta\ell)&=& v\;e^{\delta\ell}\langle e^{i G u_0^>(z)}\rangle_0 ,
\nonumber\\
&=& v\; e^{\delta\ell} e^{-{1\over2}G^2\langle 
(u_0^>(z))^2\rangle_0},\nonumber\\
&\equiv& v\;e^{(1-\eta/2)\delta\ell}
\label{vell}
\end{eqnarray}
where $\eta$ is defined by root-mean-squared phonon fluctuations
\begin{eqnarray}
\eta &\equiv& {G^2\over\delta\ell}\langle (u_0^>(z))^2\rangle_0,\nonumber\\
&=&{G^2\over 2\pi\delta\ell}{k_B T\over\sqrt{K B}}
\int_{G_z e^{-\delta\ell}}^{G_z}{d q_z\over|q_z|},\nonumber\\
&=&{G^2\over 2\pi}{k_B T\over\sqrt{K B}}\nonumber\\
&=&{k_B T\over \pi\epsilon_0},
\label{eta}
\end{eqnarray}
Because of the aforementioned relation of our model to a variety of
other problems in condensed matter physics, this RG analysis is in
fact quite familiar from those other
contexts\cite{KaneFisher,RSJ,1dIsing,Kondo}. As advertised it clearly
predicts a phase transition at $\eta=2$,\cite{Affleck} corresponding
to
\begin{eqnarray}
k_B T_p &=& 4\pi G^{-2}\sqrt{K B},\\
&=&2\pi\epsilon_0,\nonumber\\
\label{Tp}
\end{eqnarray}
between the pinned phase for $T<T_p$ and the depinned phase for
$T>T_p$. The two phases are distinguished by the relevance and
irrelevance of the pinning potential $v$, respectively. More
physically, for $T>T_p$ thermal fluctuations of the vortex lattice at
long scales effectively average away the effects of the pinning
potential (but see below), reducing it relative to the elastic energy.

In contrast for low temperatures $T<T_p$, as indicated by the RG flow,
Eq.\ref{vell}, the strength of the pinning potential, no matter how
weak at the lattice scale $a_z$ grows {\em relative} to and becomes
stronger than the typical elastic energy.  Quite clearly, since the
computation of the RG flow in Eq.\ref{vell} is done perturbatively in
$v$ the flow terminates on scale $\xi(T)\equiv a_z e^{\ell_*}$ when
elastic and pinning energies become comparable. On scales longer than
$\xi(T)$, the displacement field is well-localized at a minimum of the
periodic pinning potential, which therefore can be safely
Taylor-expanded to quadratic order and treated as a ``mass'' for
$u_0(z)$. Hence $\xi$ is determined by the balance of the effective
pinning and elastic energies on scale $\xi$
\begin{eqnarray}
{2\sqrt{B K} \over \xi}\left({a\over 2\pi}\right)^2 = v
\left({\xi\over a_z}\right)^{-\eta/2},
\label{xi_find}
\end{eqnarray}
with the effective pinning strength $v_{\rm eff} =
v(\xi/a_z)^{-\eta/2}\le v$ reduced by thermal fluctuations of modes in
the range of scales $a_z < z < \xi$. Solving Eq.\ref{xi_find} for
$\xi$ we find
\begin{eqnarray}
\xi(T) \approx \cases{\xi_0,& $\xi_0 \lesssim a\sqrt{K\over B}$\cr 
\xi_0\left({\xi_0\over a}\sqrt{B\over K}\right)^{\eta\over 2-\eta}, & 
$\xi_0 > a\sqrt{K\over B}$\cr},
\label{xi}
\end{eqnarray}
that, for weak short-scale pinning is, as expected, exponentially
lengthened by thermal fluctuations
\begin{eqnarray}
\xi(T) \approx \xi_0 e^{\sigma T/(T_p-T)},\ \ 
{\rm for}\ \ \xi_0 > a\sqrt{K\over B},
\label{xiExp}
\end{eqnarray}
with, $\sigma=\ln(\xi_0\sqrt{B/K}/a) > 0$.

We next turn to the computation of correlation functions in each of
these phases.

\subsection{Depinned phase, $T>T_p$, $h=0$}
\label{depinned_sec}

Irrelevance of the pinning potential for $T>T_p$ allows us to compute
correlation functions in the depinned phase perturbatively in $v$.  To
zeroth order, it is computed via a Gaussian integral with Hamiltonian
$\H_{el}$, Eq.\ref{Hel2d}, at long scales giving
\begin{eqnarray}
\hspace{-1cm}C_{T>T_p}(x,z)&=&\langle(u(x,z)-u(0,0))^2\rangle,\\
&\approx&2 k_B T\int{d q_x d q_z\over (2\pi)^2} {1-e^{i\q\cdot\r}\over
B q_x^2+K q_z^2},\nonumber\\
&\approx&
{k_B T\over\pi\sqrt{K B}}
\ln\left[a^{-1}\sqrt{x^2+{B\over K}z^2}\right],
\label{Cv02d}
\end{eqnarray}
where we used vortex lattice spacing $a$ as a natural short-scale
cutoff on $x$.  At the location of the defect at $x=0$,
$C_{T>T_p}(0,z)$ can be equivalently computed using the effective 1D
Hamiltonian $\H_0$, Eq.\ref{Htilde0}, to zeroth order ($v=0$) giving
\begin{eqnarray}
C_{T>T_p}(0,z)&=&\langle(u_0(z)-u_0(0))^2\rangle,\\
&\approx&{k_B T\over\pi\sqrt{K B}}\int_0^{G_z}{d q_z\over q_z} 
({1-\cos q_z z}),\nonumber\\
&\approx&{k_B T\over\pi\sqrt{K B}}
\ln\left({|z|\over a}{\sqrt{B\over K}}\right).
\label{Cv01d}
\end{eqnarray}

Vortex density correlations can be similarly computed using Eq.\ref{nv}
and phonon correlations above. To zeroth order in $v$, the average
density is given
\begin{eqnarray}
\langle n(x,z)\rangle_0 &=& n_0 - n_0\langle\partial_x u\rangle_0 +
2n_G{\rm Re}\langle e^{i2\pi n_0(x + u(x,z))}\rangle_0,\nonumber\\
&\approx& n_0 + \left({a\over L}\sqrt{K\over B}\right)^{\eta/2}
2n_G\cos(2\pi n_0 x),
\nonumber\\
&\approx& n_0,
\label{n0}
\end{eqnarray}
with fluctuations vanishing in the thermodynamic
($L\rightarrow\infty$) limit. Physically we do not expect this to be
the case, as the pinning potential breaks translational invariance in
$x$, even when it is irrelevant. This reveals that for $T>T_p$ the
pinning potential is in fact {\em dangerously} irrelevant, as dropping
it completely (as in above calculation) restores translational
invariance and gives a result that is not even qualitatively correct
for $\langle n(x,z)\rangle$. Indeed to correctly capture the behavior
of $\langle n(x,z)\rangle$, we need to compute it to at least 1st
order in $v$. The irrelevance of $v$ for $T > T_p$ guarantees the
convergence of such perturbation theory. We find\cite{Affleck}
\begin{eqnarray}
\hspace{-0.5cm}
\langle n(x,z)\rangle - n_0 
&\approx&2{n_G v\over k_B T}{\rm Re}\int dz'
\langle e^{i G(x + u(x,z))}\cos[G u(0,z')]\rangle_0,\nonumber\\
&\approx&{n_G v\over k_B T}\cos G x 
\int dz' e^{-{G^2\over2}\langle[u(x,z)-u(0,z')]^2\rangle_0},\nonumber\\
&\approx& c_\eta{n_G v a\over k_B T}\sqrt{K\over B}
\left({a\over|x|}\right)^{\eta-1}\cos(2\pi n_0 x),\nonumber\\
&&
\label{n1}
\end{eqnarray}
($c_\eta=O(1)$ dimensionless constant) showing that even in the phase
where it is irrelevant, the pinning potential leads to Friedel
oscillations in the density, a result missed by simply setting $v=0$.
This illustrates often under-appreciated distinction between
irrelevance in the RG sense and unimportance of an operator in the
physical sense.

Density two-point correlations can also be straightforwardly
computed. Using above phonon correlations and Eq.\ref{nv}, we
find\cite{Affleck}
\begin{widetext}
\begin{eqnarray}
\langle n(x,z)n(x',0)\rangle&\approx&
\langle n(x,z)\rangle_0 \langle n(x',0)\rangle_0
+{n_0^2 k_B T\over 2\pi}\sqrt{K\over B} {B z^2 - K (x-x')^2
\over [K (x-x')^2 + B z^2]^2}
+2n_G^2\left(a^2\over (x-x')^2+{B\over K} z^2\right)^{\eta/2}
\cos(2\pi n_0 x)\nonumber\\
\label{nn}
\end{eqnarray}
\end{widetext}
with the first term given by Eq.\ref{n1}, and with the last term
approximated by its lowest order (in $v$) translationally invariant
expression.

\subsection{Pinned phase, $T < T_p$, $h=0$}
\label{pinned_sec}

In the low temperature pinned phase, $v$ is relevant, growing with
increasing length scale relative to the elastic energy. On scales
longer than $\xi$, we can therefore approximate its strong pinning
effects by replacing it by a Dirichlet boundary condition on $u(x,z)$
at the location of the pin away from its ends (for a more systematic
justification of this, see Appendix \ref{Bulk_pinned}), namely taking
\begin{equation}
u(0,z)=u_0(z)=0.
\label{Dbc}
\end{equation}
Bulk phonon and density correlations with this boundary condition can
still be computed exactly. As derived for a general case in Appendix
\ref{Bulk_pinned}, a pinned phonon correlation function can be expressed
in terms of a pin-free correlation function, i.e., from a purely
harmonic elastic theory with $v=0$:
\begin{eqnarray}
G_{T<T_p}(x,x';z)&=&\langle u(x,z) u(x',0)\rangle,\\
&=&G_0(|x-x'|,z)-G_0(|x|+|x'|,z).\nonumber\\
&&
\end{eqnarray}
where $G_0(x,z)$ is $v=0$ two-point phonon correlation function. As
expected, because of the pin at $x=0$, $G_{T<T_p}(x,x';z)$ is clearly
not translationally invariant in $x$, depending on both $x$ and $x'$.
It is easy to see that it vanishes identically for $x$ and $x'$ on
opposite sides of the pin, e.g., $x>0$ and $x'<0$, showing that
because of the pin such phonon fluctuations are completely
uncorrelated on length scales longer than $\xi$. On the other hand,
for $x$ and $x'$ on the same side of the pin (e.g., $x>0$ and $x'>0$),
above expression is given by
\begin{widetext}
\begin{eqnarray}
G_{T<T_p}(x,x';z)&=&G_0(|x-x'|,z)-G_0(x+x',z),\;\;{\rm for}\;\; x>0,\; x'>0,
\nonumber\\
&=&{1\over2}\langle[u(x,z)-u(-x',0)]^2\rangle_0-
{1\over2}\langle[u(x,z)-u(x',0)]^2\rangle_0,\\
&=&{k_B T\over4\pi\sqrt{K B}}\ln\left[{K(x+x')^2+ B z^2\over
K (x-x')^2+ B z^2}\right],
\label{Gpinned}
\end{eqnarray}
\end{widetext}
and at the same point ($x=x'$, $z=z'$), reduces to
\begin{eqnarray}
\hspace{-1cm}
G_{T<T_p}(x,x;0)&=&G_0(0,0)-G_0(2x,0),\\
&=&{1\over2}\langle[u(x,0)-u(-x,0)]^2\rangle_0,\\ &\approx&{k_B
T\over2\pi\sqrt{K B}}\ln\left({2 |x|\over\overline{a}}\right).
\label{Gxxpinned}
\end{eqnarray}
In above, we naturally cut off separation at short scales by
$\overline{a}\equiv{\text{Max}}[a,\xi\sqrt{B/K}]$, since the Dirichlet
boundary condition, Eq.\ref{Dbc} is only valid on scales longer than
$\overline{a}$.  Utilizing this result, we can easily compute the
average vortex density at point $x,z$
\begin{eqnarray}
\hspace{-1cm}\langle n(x,z)\rangle 
&\approx& n_0 + 2n_G\left({\overline{a}\over
2|x|}\right)^{\eta/2}\cos(2\pi n_0 x).
\label{n0pinned}
\end{eqnarray}
$\langle n(x,z)\rangle$ in the pinned phase also displays Friedel
oscillations similar to that of the depinned phase,
Eq.\ref{n0},\ref{n1}, but with an amplitude that is nonperturbative in
$v$ and power-law fall-off with exponent $\eta/2$, rather than
$\eta-1$.  The latter is continuous at the roughening transition,
where $\eta(T_p)/2=\eta(T_p)-1=1$. This is in agreement with the
result first found by Affleck, et al. by utilizing an equivalent
Luttinger liquid phenomenology\cite{Affleck}.

\subsection{Finite $T$ and finite tilt-field $h$}
\label{finiteTh}

The calculations of previous sections have focused on either thermal
properties at a vanishing transverse field, or on finite transverse
field response at a vanishing temperature. As we show below, these can
be extended to a general point $T > 0$, $h>0$ on the phase diagram,
Fig.\ref{phasediagram}.

\subsubsection{Finite $T$ tilting transition}
The zero-temperature results for the tilting transition of
Sec.\ref{Hperp} can be extended to a finite temperature by utilizing
RG analysis and matching from Sec.\ref{depinningTransition}. As in a
related sine-Gordon commensurate-incommensurate
transitions\cite{Pokrovsky,Coppersmith} the basic nature of the
transition remains the same but with thermally renormalized effective
parameters. The extension of the zero-temperature transverse field
boundary, Eq.\ref{hc1} to a finite-temperature phase boundary
$h_{c1}(T)$ is determined by the difference in {\em free-energies} of
the commensurate and a single-soliton states, computed for
$T<T_p$. The former is simply given by the sum of the energy of the
commensurate state $E_C$ and the entropic free-energy contribution
$F_u=-k_B T\ln\Omega_u$ of the phonon modes.

The computation of a single-soliton free energy is more complicated,
since at a low temperature the pinning potential is relevant and
therefore requires a strong-coupling analysis. This can be done by
matching free-energy calculations on scales shorter and longer than
the strong coupling correlation length $\xi(T)$. As is clear from the
analysis leading to $\xi(T)$, Eq.\ref{xi}, for $\xi_0 \gg a_z$
(corresponding to weak pinning $v$) the pinning potential is
subdominant on scales $a_z < z < \xi(T)$ and the free-energy is
dominated by $F^>_u$ due to Gaussian phonon fluctuations.  On scales
longer than $\xi(T)$, the pinning potential dominates and free-energy
is determined by the sum of three contributions: (i) a single soliton
energy $E_1(T)$, with short-scale cutoff $\xi(T)$ replacing $\xi_0$ in
Eq.\ref{Es1dim}, (ii) soliton's positional entropic contribution $-k_B
T\ln(L/\xi(T))$, and (iii) free-energy contribution $F^<_u$ due to
phonon fluctuations about a single soliton configuration. Noting that
phonon fluctuations about the commensurate and single soliton states
are approximately the same, i.e., $F_u\approx F_u^> + F_u^<$, we find
that the free energy {\em difference} between a single soliton and the
commensurate state is given by:
\begin{eqnarray}
\hspace{-1cm}F_{s1}(T)&\approx&E_{s1}(T)-k_B T\ln{L\over\xi(T)},\nonumber\\
&\approx&\epsilon_0\left[2\pi\ln{L\over\xi(T)}-4\pi{L\over a} h\right]
-k_B T\ln{L\over\xi(T)},\nonumber\\
&\approx& T_p{2L\over a}\left[{a\over 2L }\left(1-{T\over T_p}\right)
\ln{L\over\xi(T)} - h\right].
\label{Fs1dim}
\end{eqnarray}
This leads to the prediction for $h_{c1}(T)$ quoted in the
Introduction:
\begin{eqnarray}
h_{c1}(T)&\approx& {a\over 2L}\left(1-{T\over T_p}\right)\ln{L\over\xi(T)},\\
&\approx& h_{c1}(0)\left(1-{T\over T_p}\right).
\label{hc1T}
\end{eqnarray}
In going from the first to second line we used a
temperature-independent approximation $\xi_0$ for $\xi(T)$,
Eq.\ref{xi}, valid for large $L$.

\subsubsection{Correlations at $T>0$, $h>0$}

Correlation functions at a finite $h$ and $T$ can also be
computed. The nature of valid approximation and their resulting form
strongly depends on three possible regimes on the phase diagram,
Fig.\ref{phasediagram}.

(a) {\em Incommensurate (fully-tilted) state, $h > h_{c2}(T)$:}

The strongly incommensurate, high tilt and temperature regime $h >
h_{c2}(T)$, where the effects of the pin can be treated perturbatively
is simplest to analyze, as was first done in
Ref.~\onlinecite{Affleck}. This can be done in close analogy to
Sec.\ref{depinned_sec}, extending it to a finite $h$. The governing
Hamiltonian is given by
\begin{eqnarray}
\H&=&{1\over2}\int dx dz\left[K(\partial_z\tilde{u})^2 + B(\partial_x
\tilde{u})^2\right]\nonumber\\
&&- v \int dz \cos[G(\tilde{u}(0,z)+h z)],
\label{Htilde2}
\end{eqnarray}
with $\tilde{u}(x,z)$ fluctuating around $0$ and to zeroth order in
the pin strength $v$ exhibiting translationally invariant
correlations:
\begin{eqnarray}
\langle(\tilde{u}(x,z)-\tilde{u}(0,0))^2\rangle_0
&\approx& {k_B T\over\pi\sqrt{K B}}
\ln\left[a^{-1}\sqrt{x^2+{B\over K}z^2}\right].\nonumber\\
\label{uutilde}
\end{eqnarray}
To the same zeroth order the average vortex density is uniform,
$n_0$. To compute it to the lowest nontrivial order in $v$, we expand
the Boltzmann weight to first order in the pinning potential, $v$ and
utilize the phonon correlations of $\tilde{u}(x,z)$, Eq.\ref{uutilde}
about the tilted state. This leads to:
\begin{widetext}
\begin{eqnarray}
\hspace{-0.5cm}
\langle n(x,z)\rangle - n_0&\approx&
{n_G v\over k_B T}{\rm Re}\left[e^{i G x}\int dz'
e^{-{G^2\over2}\langle[\tilde{u}(x,z)-\tilde{u}(0,z')]^2\rangle_0}
e^{i G h (z-z')}\right],\nonumber\\
&\approx&{n_G v\over k_B T}{\rm Re}\left[e^{i G x}\int dz'
\left({a\over\sqrt{x^2+{B\over K}(z-z')^2}}\right)^{\eta}
e^{i G h (z-z')}\right],\nonumber\\
&\approx& {n_G v a\over k_B T}\sqrt{K\over B}g_\eta(|x|/a)
e^{-|x|/\lambda_h^\infty}
\cos(2\pi n_0 x),\nonumber\\
&&
\label{n1b}
\end{eqnarray}
\end{widetext}
with the transverse field $h$ giving an exponential fall off of
Friedel oscillations on scale $\lambda_h^\infty$,
Eq.\ref{lambda_hinfty},\cite{Affleck} and preexponential function
$g_\eta(|x|/a)$ approximated by a power-law
\begin{eqnarray}
g_\eta(|x|/a)&\approx& 
\cases{c_\eta\left({a\over |x|}\right)^{\eta-1},
& $a\ll|x|\ll \lambda_h^\infty$,\cr 
d_\eta\left({a\over\lambda_h^\infty}\right)^{\eta/2-1}
\left({a\over 2|x|}\right)^{\eta/2},& $\lambda_h^\infty\ll |x|$,\cr}\nonumber\\
\label{geta}
\end{eqnarray}
$c_\eta$, $d_\eta$ O(1) dimensionless constants.

The exponential decay of density correlations, Eq.\ref{n1b} vindicates
a perturbative treatment (in $v$) at large transverse field and
temperature, where the decay length $\lambda_h^\infty$ is the shortest
scale in the problem.

(b) {\em Commensurate (pinned) transverse Meissner state, $h <
h_{c1}(T)$:}

Vortex correlations are also simple to analyze in the low tilt-field,
low-temperature commensurate $h<h_{c1}$ state, where, because of the
commensurate-incommensurate (tilting/roughening) transition we expect
a behavior that is qualitatively different from that of the
incommensurate state. We expand the Hamiltonian, Eq.\ref{Htilde2} in
vortex lattice displacements $w(x,z)$ about the nontrivial
zero-temperature distortion $u_0(x,z)$, Eq.\ref{uxz2a} (with $n_s=0$)
characterizing this state:
\begin{equation}
\tilde{u}(x,z)=\tilde{u}_0(x,z)+w(x,z).
\end{equation}
To quadratic order in $w$ we find
\begin{eqnarray}
\delta\H &=&\int dx dz\left[{K\over2}(\partial_z w)^2 +
{B\over2}(\partial_x w)^2\right]\nonumber\\
&+& {1\over 2}v\ (2\pi n_0)^2\int dz w(0,z)^2,
\label{deltaH}
\end{eqnarray}
where we used the fact that in the commensurate state $u_0(0,z)=0$.
Clearly then, because the pinning nonlinearity is localized at $x=0$,
vortex correlations of $w(x,z)$ on top of the ground state background
$u_0(x,z)$ are identical to those of the $h=0$ state. Namely, on
scales longer than the pinning length $\xi$, $w(x,z)$ fluctuations are
that of a field that is harmonic in the bulk, but pinned at the $x=0$
boundary. This identification allows us to take over results from
Sec.\ref{pinned_sec}. Using the simplified version for $u_0(x,z)$,
Eq.\ref{uxz2b}, we therefore find the average vortex density
\begin{widetext}
\begin{eqnarray}
\langle n(x,z)\rangle &=& n_0 - n_0\partial_x u_0(x,z) + 
2 n_G {\rm Re}\left[e^{i G(x + u_0(x,z))}
\langle e^{i G w(x,z)}\rangle\right],\\
&\approx& n_0 - n_0\sqrt{K\over B}{\pi\over L} h z\ {\rm sgn}(x)
e^{-\sqrt{K\over B}{\pi\over L}|x|} + 
2 n_G\left({\overline{a}\over
2|x|}\right)^{\eta/2}\cos\left[2\pi n_0(x+u_0(x,z))\right],
\label{nThlesshc1}
\end{eqnarray}
\end{widetext}
that exhibits power-law decaying Friedel oscillations
(non-perturbative in $v$) as well as a smooth compression and dilation
of the vortex lattice around the pin for $z\neq 0$. Both features can
be clearly seen in the vortex configuration displayed in
Fig.\ref{soliton0Fig}.

(c) {\em Incommensurate (soliton-tilted) state, $h_{c1}(T) < h \ll
h_{c2}(T)$:}

\begin{figure}[bth]
\centering
\setlength{\unitlength}{1mm}
\begin{picture}(40,60)(0,0)
\put(-25,0){\begin{picture}(30,40)(0,0)
\includegraphics{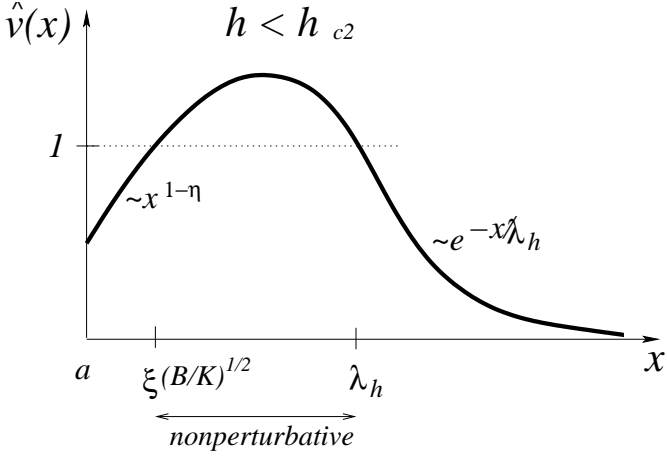}
\end{picture}}
%\put(-20,40) {$n_p$}
%\put(60,2) {$\epsilon_p$}
\end{picture}
\caption{A schematic of the length-scale dependent pinning coupling
$\hat{v}(x)$, indicating its intermediate nonperturbative regime for $h <
h_{c2}(T)$.}
\label{vxFig}
\end{figure}

We now turn to the intermediate incommensurate regime of $h$ and $T$,
with $h>h_{c1}(T)$.  Because there is only a single CI transition
(that across $h_{c1}$, with $h_{c2}$ simply a crossover), we do not
expect any {\em qualitative} change in correlations from those found
above for large $h$ and $T$. However, examining the expression for
$\langle n(x,z)\rangle$, Eqs.\ref{n1b}, \ref{geta}, the prefactor
$(a/\lambda_h^\infty)^{\eta/2-1}$, that is large for $\eta < 2$ and
$\lambda_h^\infty\gg a$, suggests a failure of the perturbative
treatment for $T<T_p$ and for a sufficiently low $h$. Indeed, as
illustrated in Fig.\ref{vxFig}, in such a regime
$\lambda_h^\infty$ will exceed the pinning length
$\xi\sqrt{B/K}$. Since for $T<T_p$, on scales longer than $\xi
\sqrt{B/K}$, even a (bare) weak pinning potential becomes comparable to the 
elastic energy and therefore cannot be treated
perturbatively. Consequently, for $\xi\sqrt{B/K}<\lambda_h^\infty$
there opens up an intermediate regime,
$\xi\sqrt{B/K}<x<\lambda_h^\infty$, where above perturbative (in $v$)
analysis leading to exponential decay of Friedel oscillations,
Eqs.\ref{n1b}, \ref{geta} fails. The crossover boundary $h_*(T)$
separating the perturbative and nonperturbative regimes is clearly
given by $\xi\sqrt{B/K}=\lambda_h^\infty$, which (not surprisingly,
but reassuringly) is equivalent to $h_*(T)\approx h_{c2}(T)\approx
a/2\pi\xi(T)$ found in Eq.\ref{hc2}.

As schematically illustrated in Fig.\ref{vxFig}, in the parlance of
RG, for $T<T_p$ the dimensionless pinning coupling $\hat{v}(x)$ at
short scale $x$ grows as $x^{1-\eta/2}$. For $h > h_{c2}(T)$, {\em
before} $\hat{v}(x)$ reaches a nonperturbative regime (i.e., reaches
$O(1)$), this growth is ``quenched'' by the tilt field $h$ on scale
longer than $\lambda_h^\infty$, beyond which $\hat{v}(x)\sim
e^{-x/\lambda_h^\infty}$ decreases exponentially, validating the fully
perturbative treatment of case (a) above. In contrast, for $h <
h_{c2}(T)$, $\hat{v}(x)$ becomes large on scale $\xi\sqrt{B/K} <
\lambda_h^\infty$, requiring a nonperturbative analysis 
on longer scales.

Above discussion clearly indicates a need for separate treatments of
three distinct regimes of length scales (i) $a<x<\xi\sqrt{B/K}$, (ii)
$\xi\sqrt{B/K}<x<\lambda_h$, (iii) $\lambda_h < x$. In the regime (i)
of shortest lengths scales, the pinning potential in $\H$,
Eq.\ref{Htilde2} can be treated perturbatively, with average density
given by the $x\ll\lambda_h^\infty$ limit of the result in
Eq.\ref{n1b}, with exponential factor approximately $1$. On longer
scales, in regime (ii) the pinning potential is comparable to the
elastic energy and vortex configurations displays a soliton array of
density $n_s(h)$, illustrated in Fig.\ref{soliton3Fig}. In this regime
$\xi\sqrt{B/K}< x <
\lambda_h^\infty <\lambda_h$ segments of vortex lines on the corresponding 
scales shorter than $n_s^{-1}$ appear to be strongly pinned in the
vicinity of the columnar defect. Hence on these scales we expect
vortex correlations of the transverse Meissner state computed in
Eq.\ref{nThlesshc1}, with only small corrections from soliton array
fluctuations. Finally, on scales $z > n_s^{-1}$ (corresponding to $x >
\lambda_h$ of regime (iii)), it is
clear from Fig.\ref{soliton3Fig}, that vortex lattice displays an
average tilt $a n_s(h)$,
\begin{equation}
u(x,z)=a n_s z + \w(x,z),
\end{equation}
with small fluctuations $\w(x,z)$ about the soliton state arising from
vibration of the soliton array. We expect the corresponding effective
Hamiltonian for $\w(x,z)$ to be given by
\begin{eqnarray}
\delta\overline{\H}
&=&{1\over2}\int dx dz\left[\K(\partial_z \w)^2 + \B(\partial_x
\w)^2\right]\nonumber\\ 
&& - \v \int dz \cos\big(G(\w(0,z)+ a n_s z)\big),
\label{Hw}
\end{eqnarray}
with $\K\approx K$, $\B\approx B$ the effective moduli of the
soliton-tilted vortex lattice and $\v\approx v(\xi/a_z)^{-\eta/2}$ a
weak pinning potential, reduced by thermal fluctuations from scales
$a_z < z < \xi$ of regime (ii). On longer scales $z > n_s^{-1}$
(regime (iii)), the soliton spacing $n_s^{-1}$ is the shortest scale
in the problem (beyond the lattice spacing), and analysis of $w$
correlations using $\delta\overline{\H}$ can be done perturbatively in
$\v$ similar to that of regime (a), that led to Eq.\ref{n1b}. As can
be seen from comparing Hamiltonians in Eq.\ref{Htilde2} and
Eq.\ref{Hw}, the main qualitative difference is the replacement of $h$
by $a n_s(h)$ inside the pinning potential, that leads to the
corresponding replacement of $\lambda_h^\infty(h)$ by the longer
length $\lambda_h(h)$ in the exponential decay of Friedel oscillations
of the average vortex density. Putting these results together for
$h_{c1} < h < h_{c2}$ we predict
\begin{widetext}
\begin{eqnarray}
\langle n(x,z)\rangle - n_0&\sim&\cos(2\pi n_0 x)
\cases{{n_G v a\over k_B T}\sqrt{K\over B}\left({a\over |x|}\right)^{\eta-1},
& $a < |x| < \xi\sqrt{B/K}$,\cr
2n_G\left({\a\over 2|x|}\right)^{\eta/2},
& $\xi\sqrt{B/K} < |x| < \lambda_h$,\cr
{n_G \v \lambda_h\over k_B T}\sqrt{K\over B}
\left({\lambda_h\over |x|}\right)^{\eta-1}e^{-|x|/\lambda_h},
& $\lambda_h < |x|$.\cr}\nonumber\\
\label{n1c}
\end{eqnarray}
\end{widetext}

\section{Finite density of independent columnar defects}
\label{finiteDefects}

So far we have focused on an idealized problem of a single columnar
defect.  As discussed in the Introduction, no genuine tilting CI phase
transition is possible in this case, since pinning energy density and
the associated lower critical field $h_{c1}$, Eq.\ref{hc1} vanish in
the 2D bulk thermodynamic limit. In this section we extend our results
to a physically more interesting case of a finite dilute concentration
$1/d$ of columnar defects.

A full treatment of such a highly nontrivial problem is beyond the
scope of the present paper and has been a subject of numerous
studies.\cite{FisherLee,NelsonVinokur,Hwa,BalentsEPL,NelsonRadzihovsky,Wallin}
Here we will be content with an approximate analysis of a dilute
concentration of columnar defects on intermediate length scales, where
vortex lattice response around each pin can be treated
independently. However, we expect that for a fixed defect
concentration (even if dilute), the system will crossover to the
anisotropic (Bose) vortex glass collective-pinning phenomenology and
our results will break down on sufficiently long scales. This is in
the spirit of other problems in physics, most notably the Kondo
effect\cite{Kondo}, where for a sufficiently dilute concentration of
impurities, on intermediate scales local moments can be treated
independently, but may order magnetically at sufficiently long scales
and low temperatures.

The main qualitative effect of a finite pin spacing $d$ should be
clear from the analysis of Sec.\ref{bulk_distortion}, in particular
from Eqs.\ref{uxq},\ \ref{uxz2a} and vortex configuration illustrated
in Fig.\ref{soliton3Fig}. There we have shown that a vortex lattice
distortion of wavelength $\lambda_z$ along the defect penetrates to
length $\lambda_\perp = \lambda_z\sqrt{B/K}/2\pi$ along $x$ into the
bulk, away from the defect. For a finite concentration of columnar
defects, we therefore expect long wavelength elastic distortions with
$\lambda_z > 2\pi d\sqrt{K/B}$, corresponding to overlapping distortion
clouds of neighboring pins to be cut off at scale $d$ along $x$.

To demonstrate this in detail, we need to generalize our results in
Eqs.\ref{uxq},\ \ref{uxz2a} to a nontrivial boundary condition on the
elastic distortion $u(x,z)$ at $x=d$, and use it to recalculate
predictions of previous sections, most importantly the energies of the
commensurate (aligned) and incommensurate (soliton) states.  The
correct qualitative physics that we are after here can be obtained by
the use of periodic boundary condition on $u(x,z)$ with period $d$, or
even more simply (and a bit cruder) a boundary condition of simply
cutting off $x$ integrals beyond length $d/2$ around each pin,
corresponding to the Dirichlet boundary condition on $u(x,z)$.

To see this in more detail, we recalculate the energy per columnar
defect $E=\H[u_0(x,z)]$, Eq.\ref{H},
\begin{eqnarray}
E&=&{1\over2}\int_0^L dz\int_{-d/2}^{d/2}dx\left[K(\partial_z u_0 -
h)^2 + B(\partial_x u_0)^2\right]\nonumber\\ 
&& - v \int_0^L dz \cos\big(G u_0(0,z)\big),
\label{Ebc1}
\end{eqnarray}
with the above boundary condition at $|x|=d/2$ and $u_0(x,z)$ given by
Eq.\ref{uxz2a}. Since Euler-Lagrange equation ensures that
contributions from each of the three terms balance each other,
$E$ approximately reduces to
\begin{eqnarray}
E&\approx&\int_0^L dz\int_{0}^{d/2}dx K\left[\partial_z\overline{u}_{0}(x,z) -
h_{\rm eff}e^{-\sqrt{K\over B}{\pi\over L}|x|}\right]^2-v L.\nonumber\\ 
\label{Ebc2}
\end{eqnarray}
Calculating above expression we find, that, for $L < 2\pi d\sqrt{K/B}$
it reduces to the previously calculated single-pin result given in
Eq.\ref{ENs}. In the opposite regime, $L > 2\pi d\sqrt{K/B}$ of
interest to us, $E$ crosses over to
\begin{eqnarray}
E(d)&\approx&E_C(d)+ K a d(h_{c1}-h) N_s +{1\over2} K a^2 {d\over
L}N_s^2,
\nonumber\\
\label{Ebc3}
\end{eqnarray}
with bulk energy per pin
\begin{eqnarray}
E_C(d)&=& K h^2 L d - v L,
\label{ECd}
\end{eqnarray}
leading to 
\begin{eqnarray}
h_c(d)&=&\sqrt{v\over K d},
\label{hcd}
\end{eqnarray}
and
\begin{eqnarray}
h_{c1}(d)&\approx&\sqrt{B\over K}{a\over 4\pi d}
\ln\left[\sqrt{K\over B}{2\pi d\over\xi}\right],
\label{hc1d}
\end{eqnarray}
as quoted in the Introduction.

\section{Conclusions}
\label{conclusion}

In this paper we studied a finite-temperature response of a planar
vortex array to an in-plane tilting of an external magnetic field away
from a dilute concentration of pinning columnar defects.  We found
that the vortex lattice tilting proceeds via an interesting finite
temperature, finite transverse field commensurate-incommensurate
transition at $h_{c1}(T)$, driven by a proliferation of solitons, as
illustrated in the phase diagram, Fig.\ref{phasediagram}. A sensitive
dependence of this lower-critical field (that vanishes for a single
pin) on the columnar defect spacing should be experimentally testable
by varying the heavy-ion irradiation flux used to create pinning
tracks. We show that at low-temperatures, for $h>h_{c1}(T)$ the vortex
array exhibits a highly nontrivial soliton-like distortion as a
compromise between inter-vortex interaction, and the pinning and
diamagnetic energies. We show that this nonlinear response persists up
to a large upper-critical tilting angle, $\tan\theta_{c2}\sim
a/\xi(T)$, beyond which system recovers a full linear transverse
susceptibility with $B_\perp\approx H_\perp$. We expect that these and
many other detailed predictions should be directly testable in
computer simulations. Although much more difficult, it is our hope
that the theory presented here can be furthermore tested in mesoscopic
samples of artificially layered superconductors.

More importantly, we expect that a number of these features will carry
over to a transverse field response in bulk superconductors. Extending
our two-dimensional planar results and exploring their impact on
phenomenology of bulk samples remains an important and challenging
problem.

\begin{acknowledgments}
  It is a pleasure to acknowledge illuminating discussions with M.
  Ablowitz, I. Affleck, V. Gurarie, D. R. Nelson, A. Polkovnikov, and
  J. Toner. I thank an anonymous referee for a careful reading of the
  manuscript, making valuable suggestions, and catching a number of
  typographical errors.  This research was support by the National
  Science Foundation through MRSEC DMR-0213918, DMR-0321848, and by
  the David and Lucile Packard Foundation.
\end{acknowledgments}

\appendix
\section{Boundary Hamiltonian via a functional integral}
\label{integralBC}

Problems where it is possible and convenient to eliminate bulk degrees
of freedom, thereby reducing the problem to a lower-dimensional one
are quite common in physics. In this appendix, for completeness, we
present a general functional-integrals approach to such problems, and
apply it to the problem of a planar vortex lattice pinned by a single
columnar defect treated my more pedestrian methods in the main text.

To this end, we consider a field $\phi({\bf x},{\bf z})$ defined on a
coordinate space ${\bf r}=({\bf x},{\bf z})$ that we split into bulk
${\bf x}$ and boundary ${\bf z}$ subspaces. The energetics is governed
by a Hamiltonian $\H[\phi]$ and the partition function is given by a
standard functional integral over the field $\phi({\bf r})$ ($k_B
T\equiv 1$):
\begin{equation}
Z=\int[d\phi({\bf r})]e^{-\H[\phi]}
\end{equation}
$Z$ can be equivalently expressed as an integral over field $\phi({\bf
r})$ constrained on the boundary ${\bf x}=0$ to be $\phi({\bf 0},{\bf
z})=\phi_0({\bf z})$, followed by an integral over the boundary fields
$\phi_0({\bf z})$. Explicitly, the former is implemented by a
functional $\delta$-function, $\delta[\phi({\bf z})]\equiv\prod_{\bf
z}\delta(\phi({\bf z}))$ giving
\begin{eqnarray}
Z&=&\int[d\phi_0({\bf z}) d\phi({\bf x},{\bf z})]\;
\delta[\phi({\bf 0},{\bf z})-\phi_0({\bf z})]\;e^{-\H[\phi]},\hspace{1cm}\\
&\equiv&\int[d\phi_0({\bf z})]\;e^{-\H_0[\phi_0({\bf z})]}.
\end{eqnarray}
The effective boundary Hamiltonian $\H_0[\phi_0(z)]$ defined above can
be expressed using the Fourier representation of the functional
$\delta$-function, leading to
\begin{eqnarray}
e^{-\H_0[\phi_0(\zz)]}&=&\int[d\phi d\lambda]\;
e^{-\H[\phi]+i\int_{\bf z}\lambda({\bf z})
(\phi({\bf 0},{\bf z})-\phi_0({\bf z}))}.\nonumber\\
\label{deltaFT}
\end{eqnarray}
For a general Hamiltonian $\H[\phi]$, above computation can only be
performed via a formal cumulant expansion. However, for a special case
of a quadratic Hamiltonian
\begin{equation}
\H[\phi]={1\over2}\int_{{\bf r},{\bf r}'}\phi({\bf r})G^{-1}({\bf
r}-{\bf r}')\phi({\bf r}'),
\label{HphiAppendix}
\end{equation}
defined by a correlation function $G({\bf r})$ (with $G^{-1}({\bf r})$
its inverse), all cumulants reduce to a power of $G({\bf r})$, i.e.,
obey Wick's theorem, equivalent to the Gaussian integral identity
\begin{equation}
\int_{-\infty}^\infty d\phi e^{-{1\over2}a^{-1}\phi^2 +
\lambda\phi}=\left({2\pi\over a}\right)^{1/2}\;e^{{1\over2} a\lambda^2}.
\label{Wick}
\end{equation}  
A Gaussian functional integrations over $\phi({\bf r})$ and
$\lambda({\bf z})$ (dropping an inconsequential $\phi_0$-independent
constant) then gives
\begin{eqnarray}
e^{-\H_0[\phi_0(\zz)]}&=&\int[d\lambda]\;
e^{-{1\over2}\int_{{\bf z},{\bf z}'}\lambda({\bf z})G({\bf 0},{\bf z}-{\bf
z}')\lambda({\bf z}')-i\int_{\bf z}\lambda({\bf z})\phi_0({\bf
z})}.\nonumber\\
&=&e^{-{1\over2}\int_{{\bf z},{\bf z}'}\phi_0({\bf z})
G^{-1}({\bf 0},{\bf z}-{\bf z}')\phi_0({\bf z}')}.
\end{eqnarray}

Applying this simple result to the (1+1)-dimensional (planar) vortex
lattice pinned by a columnar defect at $x=0$ leads to
\begin{equation}
\H_0[u_0(z)]={1\over2}\int_{z,z'}u_0(z)G^{-1}_0(z-z')u_0(z'),
\end{equation}
with 
\begin{equation}
G^{-1}_0(z)\equiv G^{-1}(0,z)
\end{equation}
an inverse of the bulk propagator $G(0,z)$ evaluated at $x=0$. In
$q_z$-Fourier space the latter is easily evaluated
\begin{eqnarray}
\tilde{G}(x=0,q_z)&=&\int{d q_x\over 2\pi}{1\over K q_z^2 + B q_x^2},\\
&=&{1\over 2\sqrt{B K}}{1\over |q_z|},
\end{eqnarray}
and leads to
\begin{eqnarray}
\tilde{G}_0^{-1}(q_z)&=&2\sqrt{B K}|q_z|,
\end{eqnarray}
thereby confirming the resulting for $\H_0[u_0(q_z)]$,
Eq.\ref{Htilde0z} obtained in Sec.\ref{Reduction1d} by a different
method.

\section{Bulk correlation functions in a ``pinned'' state}
\label{Bulk_pinned}

In this appendix, as a model of phonon correlations in the ``pinned''
vortex state, we study bulk correlation function of a Gaussian field
constrained at $\r=(0,\zz)$ by a ``massive'' boundary Hamiltonian
$\H_p[\phi(\x,0)]$. As argued in the main text, on sufficiently long
scales the boundary Hamiltonian can simply be implemented as a hard
constraint on the field to vanish at $\r=(0,\zz)$. To this end we
compute the asymptotic generating function $Z[j(\r)]$
\begin{eqnarray}
\hspace{-1cm}
Z[j(\r)]&=&\int[d\phi({\bf x},{\bf z})]\;
\delta[\phi({\bf 0},{\bf z})]\;e^{-\H[\phi]+\int_\r j(\r)\phi(\r)},
\end{eqnarray}
from which, by differentiation with respect to $j(\r)$ all $n$-point
correlation functions of $\phi(\r)$ can be obtained. As in Appendix
\ref{integralBC}, we have implemented the Dirichlet boundary condition on 
$\phi(\r)$ via a functional $\delta$-function. Representing the latter
in its (functional) Fourier form as in Eq.\ref{deltaFT}, using
harmonic Hamiltonian, Eq.\ref{HphiAppendix}, and performing a Gaussian
integral over $\phi(\r)$, we find
\begin{eqnarray}
Z[j(\r)]&=&e^{{1\over2}\int_{\q,\q'}j(\q)\Gamma[\q,\q']j(\q')}.
\label{Zj}
\end{eqnarray}
The kernel $\Gamma[\q,\q']$ is given by
\begin{widetext}
\begin{eqnarray}
\hspace{-1cm}
\Gamma[\q,\q']=(2\pi)^{d_z}\delta^{d_z}(\q_z+\q_z')
\G(\q_x,\q_z)\left[(2\pi)^{d_x}\delta^{d_x}(\q_x+\q_x')
-\G(\q_x',-\q_z)\left(\int_{\q_x''}\G(\q_x'',\q_z)\right)^{-1}\right],
\label{Gamma}
\end{eqnarray}
\end{widetext}
and leads to real-space $\phi(\r)$ 2-point correlation function
\begin{eqnarray}
G_{\rm pinned}[\x,\x';\zz]
&=&\langle\phi(\x,\zz)\phi(\x',0)\rangle_{\rm pinned},
\label{Gpinnedapp}
\end{eqnarray}
that in $\q_z$-Fourier space is given by
\begin{eqnarray}
\G_{\rm pinned}[\x,\x';\q_z]&=&\G(\x-\x';\q_z)
-{\G(\x,\q_z)\G(\x',-\q_z)\over \G(0,\q_z)},\nonumber\\
&&
\label{Gpinned2app}
\end{eqnarray}

In the special (1+1)-dimensional case of $\G^{-1}(q_x,q_z)=K q_z^2+B
q_x^2$ of interest to us in the main text,
\begin{eqnarray}
\G(x,q_z)&=&\int{d q_x\over 2\pi}{e^{i q_x x}\over K q_z^2 + B q_x^2},\\
&=&{e^{-({K\over B})^{1\over2}|q_z||x|}\over 2\sqrt{B K}|q_z|},
\label{Gxqz}
\end{eqnarray}
and leads to
\begin{widetext}
\begin{eqnarray}
G_{\rm pinned}[x,x';z]&=&{1\over 2\sqrt{B K}|q_z|}
\left[e^{-({K\over B})^{1\over2}|q_z||x-x'|}
-e^{-({K\over B})^{1\over2}|q_z|(|x|+|x'|)}\right],\\
&=&G(|x-x'|,z)-G(|x|+|x'|,z).
\label{Gpinned3app}
\end{eqnarray}
\end{widetext}
utilized in Eq.\ref{Gpinned} of the main text.

A simpler way to derive above ``pinned'' correlation function result
is to note that the boundary condition $\phi({\bf 0},{\bf z})=0$ is
automatically explicitly satisfied by the odd part of $\phi({\bf
x},{\bf z})$. It is also satisfied by a subset of (an independent)
even part of $\phi({\bf x},{\bf z})$ that vanishes at $\x={\bf 0}$.
Naively, such even/odd field decomposition does not represent a
simplification since a constraint on the even part must still be
enforced. However, it is clear that for correlations on the same side
of the pin, a constrained even part of the field has identical
correlations to that of the odd part of the field.  Hence, $G_{\rm
pinned}[\x,\x';\zz]$ for $\x \x' > 0$ (i.e., on the same side of the
pin) is simply given by twice the correlator of the unconstrained odd
part, $\phi({\bf x},{\bf z})\rightarrow {1\over2}(\phi({\bf x},{\bf
z})-\phi(-{\bf x},{\bf z}))$
\begin{widetext}
\begin{eqnarray}
G_{\rm pinned}[\x,\x';\zz]
&=&{1\over2}\langle(\phi(\x,\zz)-\phi(-\x,\zz))
(\phi(\x',0)-\phi(-\x',0))\rangle_0,\\
&=&\langle\phi(\x,\zz)\phi(\x',0)\rangle_0
-\langle\phi(\x,\zz)\phi(-\x,0)\rangle_0,
\label{GpinnedApp_again}
\end{eqnarray}
\end{widetext}
giving the result in Eq.\ref{Gpinned3app}.

We conclude this appendix with a computation of the generating
function $Z[j(\r)]$, that extends above analysis to scales shorter
than the pinning length. To this end, we supplement the harmonic bulk
Hamiltonian with a pinning one at the boundary, $\H_p[\phi(0,\zz)]$
with the full $\H$ given by
\begin{equation}
\H[\phi]={1\over2}\int_{{\bf r},{\bf r}'}\phi({\bf r})G^{-1}({\bf
r}-{\bf r}')\phi({\bf r}') + \int_{\bf r}\delta^{d_x}(\x)\H_p[\phi(0,\zz)],
\label{Hphi_bAppendix}
\end{equation}
The corresponding generating function is then given by
\begin{widetext}
\begin{eqnarray}
Z[j(\r)]&=&\int[d\phi_0({\bf z})]e^{-\H_p[\phi_0(\zz)]} 
\int [d\phi({\bf x},{\bf z})]\;
\delta[\phi({\bf 0},{\bf z})-\phi_0({\bf z})]\;
e^{-{1\over2}\int_{{\bf r},{\bf r}'}\phi({\bf r})G^{-1}({\bf
r}-{\bf r}')\phi({\bf r}')+\int_\r j(\r)\phi(\r)}.
\label{ZjApp}
\end{eqnarray}
\end{widetext}
As in Appendix \ref{integralBC}, representing the functional
$\delta$-function in its Fourier form, integrating over the harmonic
bulk and boundary Fourier fields $\phi(\r)$, $\lambda(\zz)$, we find
\begin{eqnarray}
Z[j(\r)]&=&\int[d\phi_0({\bf z})]e^{-\W[\phi_0(\zz),j(\r)]},
\label{ZjApp2}
\end{eqnarray}
with
\begin{widetext}
\begin{eqnarray}
\W[\phi_0,j]&=&H_p[\phi_0(\zz)]
-{1\over2}\int_{{\bf r},{\bf r}'}j({\bf r})G({\bf r}-{\bf r}')j({\bf r}')\\
&+&{1\over2}\int_{{\bf z},{\bf z}'}
\left[\phi_0(\zz)-\int_{\zz_1,\x_1}G_0(\x_1,\zz-\zz_1)j(\x_1,\zz_1)\right]
G^{-1}(0,\zz-\zz')
\left[\phi_0(\zz')-\int_{\zz_2,\x_2}G_0(\x_2,\zz'-\zz_2)j(\x_2,\zz_2)\right].
\nonumber
\end{eqnarray}
\end{widetext}
For $j({\bf r})$ and $\H_p[\phi_0(\zz)]=-v\int_z\cos\phi_0(z)$,
$\W[\phi_0(\zz),0]$ simply reduces to the Hamiltonian,
Eq.\ref{Htilde0} localized on the pin, with the integrated out bulk
degrees of freedom reflected in its long-range elasticity. On the
other hand, for a finite $j({\bf r})$, but strong pinning, on
sufficiently long scales $e^{-\H_p[\phi_0(\zz)]}$ simply acts as a
hard constraint $\phi_0(\zz)=0$, reducing $Z[j(\r)]$ to the previously
found result given in Eqs.\ref{Zj},\ \ref{Gamma},\ \ref{Gpinned2app}.

In the pinned (commensurate) phase, a columnar defect pins a single
vortex line, corresponding to a ``confinement'' of field $\phi_0(\zz)$
to a single minimum of the cosine and allowing us to approximate
$\H_p[\phi_0(\zz)]=-v\int_\zz\cos\phi_0(\zz)\approx {\rm const.} +
{1\over 2} v \int_\zz \phi_0(\zz)^2$ by a harmonic ``spring''. We can
therefore integrate over $\phi_0(\zz)$ in Eq.\ref{ZjApp2}, obtaining (up to
an unimportant multiplicative constant):
\begin{eqnarray}
Z[j(\r)]&=&e^{{1\over2}\int_{\r,\r'}j(\r)\Gamma_{\rm pinned}[\r,\r']j(\r')}.
\end{eqnarray}
with
\begin{widetext}
\begin{eqnarray}
\Gamma_{\rm pinned}[\r,\r']&=&
\int_{\zz_1,\zz_1'}G(\x,\zz_1'-\zz)G^{-1}(0,\zz_1-\zz_1')
\left[v\delta(\zz_1-\zz_2)+G^{-1}(0,\zz_1-\zz_2)\right]^{-1}
G(\x',\zz_2'-\zz')G^{-1}(0,\zz_2-\zz_2')\nonumber\\
&+&G({\bf r}-{\bf r}')
-\int_{\zz_1,\zz_2}
G(\x,\zz-\zz_1)G^{-1}(0,\zz_1-\zz_2)G(\x',\zz'-\zz_2).
\end{eqnarray}
\end{widetext}
In Fourier space, this becomes
\begin{widetext}
\begin{eqnarray}
\Gamma_{\rm pinned}[\q,\q']&=&(2\pi)^{d_z}\delta^{d_z}(\q_z+\q_z')
\Big[(2\pi)^{d_x}\delta^{d_x}(\q_x+\q_x')\G(\q_x,\q_z)\nonumber\\
&-&\G(\q_x,\q_z)\G^{-1}(\x=0,-\q_z)\G(\q_x',-\q_z)
{v\over v+\G^{-1}(\x=0,\q_z)}\Big].
\end{eqnarray}
\end{widetext}
For a system that is translationally invariant along $\zz$,
$\G^{-1}(\x=0,\q_z)$ generically vanishes at long wavelengths
$\q_z\rightarrow 0$, and the result reduces to that of a hard
constraint, given in
Eqs.\ref{Zj},\ \ref{Gamma}. In more detail for the (1+1)D vortex problem at 
hand, $\G^{-1}(\x=0,\q_z)=({a/2\pi})^2 2\sqrt{K B}|q_z|$, showing 
that, as asserted in the main text, the crossover to the hard
constraint happens on scales longer than the pinning length, $q_z^{-1}
\gg \xi$.

\section{Hilbert transform basics}
\label{HilbertBasics}

In this appendix, for completeness we summarize some of the basics of
Hilbert transforms necessary to derive results in the main text and in
the Appendixes.

Hilbert transform $\tilde{\phi}(y)={\rm H}[\phi(x)]$ of a function
$\phi(x)$ is defined by
\begin{equation}
{\rm H}[\phi(x)]={1\over\pi}{\rm P}\int_{-\infty}^\infty d x
{\phi(x)\over x-y},
\end{equation}
where ${\rm P}$ stands for the principal value of the integral i.e.,
with the singular point $x=y$ excluded. 

Hilbert transforms of standard functions can be usually computed by
relating it to a contour integral in a complex plane. For example, 
Hilbert transform of $\sin x$ and $\cos x$ can be computed as real and
imaginary parts of Hilbert transform of $e^{i x}$
\begin{eqnarray}
{\rm H}[e^{i x}]&=&
{1\over\pi}{\rm P}\int_{-\infty}^\infty d x {e^{i x}\over x-y},\\
&=& i e^{i y},
\end{eqnarray}
with last expression obtained easily by contour integration, taking
advantage of analyticity of $e^{i x}$ in the upper-half plane. Above
result then leads to 
\begin{eqnarray}
{\rm H}[\sin x]&=& \cos y,\\
{\rm H}[\cos x]&=& -\sin y.
\end{eqnarray}

More importantly for the problem of the vortex lattice at hand, we
compute the Hilbert transform of a Lorentzian 
\begin{eqnarray}
{\rm H}[{1\over x^2 + 1}]&=&
{1\over\pi}{\rm P}\int_{-\infty}^\infty d x {1\over x-y}{1\over x^2 + 1},\\
&=&{-y\over y^2 +1},
\label{Hneeded}
\end{eqnarray}
by noting that it is related a semi-circular contour integral over $C$
in the upper-half plane. Equivalently, it can be computed as (minus) the
imaginary part of Hilbert transform of ${x-i\over x^2 + 1}={1\over x +
i}$, with the latter function analytic in the upper-half plane. As a
side benefit the real part of ${\rm H}[{1\over x + i}]$ gives 
\begin{eqnarray}
{\rm H}[{x\over x^2 + 1}]&=& {1\over y^2 +1}.
\end{eqnarray}

\section{Soliton solution of the sine-Hilbert equation}
\label{singleSoliton}

In this appendix we verify that the soliton solution
\begin{equation}
\phi_s(z)=-2{\rm ArcTan}{1\over z}
\end{equation}
indeed satisfies the Euler-Langrange integral equation
\begin{eqnarray}
{1\over\pi}\int d z' {\phi_s(z) - \phi_s(z')\over(z-z')^2} + \sin\phi_s(z)=0,
\label{eom1Appendix}
\end{eqnarray}
where from now on, all the integrals are understood in the sense of a
principal part, a physically dictated regularization.  To this end,
using a relation ${\rm P}\int {d z'\over(z-z')^2}=0$ and integrating
by parts, we note that the sine-Hilbert equation can be rewritten as
\begin{eqnarray}
-\partial_z{\rm H}[\phi_s(z')] + \sin\phi_s(z)&=&0,\\
-{\rm H}[\partial_{z'}\phi_s(z')] + \sin\phi_s(z)&=&0.
\label{eom2Appendix}
\end{eqnarray}
Now using
\begin{equation}
\partial_z\phi_s(z)={2\over z^2 +1},
\end{equation}
and Hilbert transform relation from Appendix \ref{HilbertBasics},
Eq.\ref{Hneeded}, we find
\begin{eqnarray}
{\rm H}[\partial_{z'}\phi_(z')]={-2 z\over z^2 + 1}.
\end{eqnarray}
Then calculating
\begin{eqnarray}
\sin\phi_s(z)&=&-\sin\left[2{\rm ArcTan}{1\over z}\right],\\
&=&{-2 z\over z^2 + 1},
\end{eqnarray}
shows that indeed $\phi_s(z)$ satisfies the sine-Hilbert equation,
Eq.\ref{eom2Appendix}.

\section{Single soliton energy}
\label{solitonEnergy}

In this appendix we compute the energy $\E_1[\h]=\Hh_0[\phi_s(\z)]$ of
a single soliton,
\begin{equation}
\phi_s(\z)=-2{\rm ArcTan}{1\over \z-\z_0},
\label{solitonApp}
\end{equation}
for the sine-Hilbert model, defined by a Hamiltonian
\begin{eqnarray}
\Hh_{0}&=&\Hh_0^{el} + \Hh_0^p,\nonumber\\
&=&{1\over 4\pi}\int\int d\z d\z'\left({\phi(\z)-\phi(\z') -
\hat{h}(\z-\z')\over
\z-\z'}\right)^2\nonumber\\ 
&&-\int d\z\cos\phi(\z),
\label{H0app}
\end{eqnarray}
where all integrals are implicitly understood to range over the system
size, with $-\L/2<\z<\L/2$. Expanding the square of the elastic part,
$\E_1^{el}=\Hh_0^{el}[\phi_s(\z)]$, we find (to accuracy of
${\cal{O}}(1)$ for $\L\rightarrow\infty$):
\begin{widetext}
\begin{eqnarray}
\E_1^{el}&=&{1\over 2\pi}\int\int d\z
d\z'\left[{\phi_s(\z)^2-\phi_s(\z)\phi_s(\z')\over (\z-\z')^2}
-\h{\phi_s(\z)-\phi_s(\z')\over\z-\z'}\right]+{1\over 4\pi}\h^2\L^2,\\
&=&-{1\over2}\int d\z\phi_s(\z){\rm H}[\partial_{\z}\phi_s(\z)]
-\h\int d\z{\rm H}[\phi_s(\z)]+{1\over 4\pi}\h^2\L^2.
\label{E1elApp}
\end{eqnarray}
\end{widetext}
These integrals can be computed utilizing Hilbert transforms worked
out in Appendix \ref{HilbertBasics}.  Using Eq.\ref{Hneeded}, the
first term, $\E_{1a}^{el}$ can be integrated by parts
\begin{eqnarray}
\E_{1a}^{el}
&=&-{1\over2}\int d\z\phi_s(\z){\rm H}[\partial_{\z}\phi_s(\z)],\\
&=&-2\int d\z {\z\over \z^2+1}{\rm ArcTan}{1\over \z},\\
&=&2\pi\ln{\L\over4}.
\label{E1el_aApp}
\end{eqnarray}
Less formally, this elastic contribution can be computed by going back
to the expression
\begin{eqnarray}
\E_{1a}^{el}={1\over 4\pi}\int\int d\z
d\z'\left({\phi_s(\z)-\phi_s(\z')\over \z-\z'}\right)^2,
\label{E1el_a1App}
\end{eqnarray}
and noting that because $\phi_s(\z)$ vanishes for $\z\gtrsim 1$ and
equals $2\pi$ for $\z\lesssim -1$, finite contributions to the elastic
energy arise only from regions $(\z\lesssim -1, \z'\gtrsim 1)$ and
$(\z\gtrsim 1, \z'\lesssim -1)$. To accuracy of ${\cal{O}}(1)$, this
reduces the soliton elastic energy to
\begin{eqnarray}
\E_{1a}^{el}&\approx&{1\over \pi}\int_{-\L/2}^{-1} d\z
\int_1^{\L/2} d\z' {(2\pi)^2\over (\z-\z')^2},\\
&\approx& 2\pi\ln{\L\over4},
\label{E1el_a2App}
\end{eqnarray}
in agreement with the more formal analysis above.

The second contribution to $\E_1^{el}$ in Eq.\ref{E1elApp} can also be
computed by integrating by parts and noting that ${\rm H}[\phi_s(\z)]$
(that can be explicitly computed giving ${\rm
H}[\phi_s(\z)]=-\ln\left[{\z^2+1\over(\L/2)^2+1}\right]$) vanishes at
the boundaries of the system, $\z=\pm\L/2$
\begin{eqnarray}
\E_{1b}^{el}&=&-\h\int_{-\L/2}^{\L/2} d\z {\rm H}[\phi_s(\z)],\\
&=&\h\int_{-\L/2}^{\L/2} d\z \z {\rm H}[\partial_{\z}\phi_s(\z)],\\
&=&-\h(2\L - 2\pi).
\label{E1el_bApp}
\end{eqnarray}

A single soliton pinning contribution
$\E_1^{p}=\Hh_0^{p}[\phi_s(\z)]$ is also straightforward to calculate
using solution $\phi_s(\z)$, Eq.\ref{solitonApp}. We find
\begin{eqnarray}
\E_{1}^{p}&=&-\int_{-\L/2}^{\L/2} d\z \cos\phi_s(\z),\\
&=&-\int_{-\L/2}^{\L/2} d\z\left(1-2\sin^2(\phi_s/2)\right),\\
&=&-\int_{-\L/2}^{\L/2} d\z\left(1-{2\over \z^2+1}\right),\\
&=& -\L + 2\pi.
\label{E1pApp}
\end{eqnarray}

Combining above contributions inside Eq.\ref{H0app}, we obtain the
expression for a single soliton dimensionless energy
\begin{eqnarray}
\E_{1}&=&{1\over 4\pi}\h^2\L^2 - \L 
+ 2\pi\ln{e\L\over4}-\h(2\L - 2\pi),\;\;\;\;\;\;\\
&\approx&{1\over 4\pi}\h^2\L^2 - \L + 2\pi\ln\L-2\h\L
\label{E1App}
\end{eqnarray}
used in the main text.

\section{Two-soliton energy: interaction}
\label{soliton2Energy}
In this appendix we give a few technical details for the computation
of the soliton interaction energy. For two far-separated solitons (of
interest in a dilute soliton approximation, valid for $h\ll h_{c2}$)
with $|\z_1-\z_2|\gg 1$, we can approximate the exact two-soliton
solution by a sum of two one-soliton solutions
\begin{equation}
\phi_{2s}(\z_1,\z_2)\approx\phi_s(\z_1)+\phi_s(\z_2).
\label{phi2s}
\end{equation}
The two-soliton interaction is then determined by two-soliton energy
\begin{widetext}
\begin{eqnarray}
\hspace{-1cm}\E_{2}(\z_1,\z_2)&\approx&\Hh_0[\phi_{s1}+\phi_{s2}],\\
&=&{1\over 4\pi}\int\int d\z
d\z'\left({\phi_{s1}(\z)+\phi_{s2}(\z)-\phi_{s1}(\z')-\phi_{s2}(\z')-
\h(\z-\z')\over \z-\z'}\right)^2-\int
d\z\cos[\phi_{s1}(\z)+\phi_{s2}(\z)],\\
&=&\E_C+2\E_{s1}+V_s(\z_1-\z_2),
\label{E2sDefineApp}
\end{eqnarray}
\end{widetext}
that consists of the zero-soliton contribution $\E_C$, Eq.\ref{hatEC},
two one-soliton contributions $\E_{s1}$, and soliton interaction
$V_s(\z_1-\z_2)=V_s^A(\z_1-\z_2) + V_s^B(\z_1-\z_2)$ given by
\begin{widetext}
\begin{eqnarray}
V_s^A(\z_1-\z_2)&=&{1\over\pi}\int\int d\z
d\z'{\phi_{s1}(\z)\phi_{s2}(\z)-\phi_{s1}(\z)\phi_{s2}(\z')
\over (\z-\z')^2},\\
V_s^B(\z_1-\z_2)&=&\int d\z
\left[1-\cos\phi_{s1}(\z)\cos\phi_{s2}(\z)+\sin\phi_{s1}(\z)\sin\phi_{s2}(\z)
\right]
\label{VsApp}
\end{eqnarray}
\end{widetext}

Manipulations similar to those for the computation of a single soliton
energy give
\begin{eqnarray}
\hspace{-0.3cm}V_s^A(\z_1-\z_2)&\approx&-\int d\z\phi_{s1}(\z)
{\rm H}[\partial_{\z}\phi_{s2}(\z)],\\
&\approx&-4\int d\z {\z-\z_2\over (\z-\z_2)^2+1}{\rm ArcTan}{1\over\z-\z_1},\\
&=& 4\pi\ln{\L\over 2}-2\int_{-\infty}^\infty d\z
{\ln\left[(\z-\z_2)^2+1\right]\over(\z-\z_1)^2+1},\;\;\;\;\;\;\;\;\\
&=&4\pi\ln{\L\over2}-2\pi\ln\left[(\z_1-\z_2)^2+4\right],
\label{VsAApp}
\end{eqnarray}
with a simplifying approximation above valid for $|z_{1,2}|\ll
L/2\rightarrow\infty$.  Similarly, a straightforward contour
integration gives
\begin{eqnarray}
\hspace{-0.5cm}
V_s^B(\z_1-\z_2)&=&2\int_{-\infty}^\infty d\z {(2\z-\z_1-\z_2)^2
\over\left((\z-\z_1)^2+1\right)
\left((\z-\z_2)^2+1\right)},\;\;\;\;\;\;\;\;\;\;\\
&=& 4\pi,
\label{VsBApp}
\end{eqnarray}
that, together with Eq.\ref{VsAApp} gives the soliton interaction
\begin{eqnarray}
V_s(\z)&=&2\pi\ln\left[{(\L/2)^2\over \z^2+4}\right] + 4\pi,
\label{V_sApp}
\end{eqnarray}
used in the main text.

%\end{multicols}

\begin{references}
\bibitem{noTransition} There is no transition in finite system, as 
nonanalyticity can only appear in thermodynamic limit. Given that, in
$L\rightarrow\infty$ limit the critical value $h_{c1}$ for the
dimensionless transverse tilting field is driven to $0$, one concludes
that indeed for one pin there is no phase transition. However, as in
analogous problems, e.g., Kondo\cite{Kondo}, we imagine that in a
physical system there is a finite dilute (and therefore
noninteracting) concentration of columnar pins, which will prevent
$h_{c1}$ from being driven to $0$ and will exhibit a true CI tilting
transition studied here.
\bibitem{GLelastic} Elastic moduli $K$ and $B$ can be derived 
from the Ginzburg-Landau theory and for finite-range interactions (set
by the London penetration length $\lambda$) are wavevector dependent
on scales shorter than $\lambda$, and become dispersionless only 
on scales longer than $\lambda$, i.e., for $n_0\lambda\ll 1$.
For simplicity we will ignore this inessential complication.
\bibitem{gless1} General arguments
suggest that the effective Luttinger parameter $g$ (proportional to
$1/\sqrt{K B}$) is constrained by $1$ from above, corresponding to a
universal, noninteracting fermion
limit\cite{TonksGirardeau,Lieb,Pokrovsky,Coppersmith} reached in the
dilute vortex gas near $H_{c1}$. However, at high densities, such that
vortex spacing falls below the London penetration length elastic
moduli become wavevector dependent. Furthermore, at higher densities
and lower temperature, fluctuation (Helfrich)
contribution\cite{Pokrovsky,Coppersmith} to compressional modulus
becomes subdominant. Hence in this regime the Luttinger parameter can
exceed $1$, allowing for the roughening (thermal depinning)
transition at $g=1$.\cite{Affleck}
\bibitem{commentHc} Soliton proliferation CI transition at $h=h_{c1}$ 
is closely analogous to the transition at $H_{c1}$ from the Meissner
state to the vortex state in type II superconductors. The transverse
field $h_{c2}$, at which solitons form a dense overlapping array with
$n_s\approx 1/\xi$ corresponding to a nearly perfect vortex alignment
with an external field (i.e., absence of transverse diamagnetism),
then corresponds to the upper-critical $H_{c2}$ field at which the
system asymptotes to the normal state. In this identification, the
bulk $h_c$ field corresponds to the thermodynamic $H_c$ field of a
type-II superconductor, beyond which the energy of a bulk
superconducting Meissner state exceeds that of the normal state.
\bibitem{logInteraction} We expect that in contrast to systems in
  which solitons interact via a short-range potential (as in e.g.,
  sine-Gordon model), here, because of logarithmic soliton interaction
  we expect the approximation of a multi-soliton solution by a sum of
  one-soliton solutions to be less accurate. Unfortunately, it is
  quite challenging to go beyond this approximation analytically. 
  Nevertheless, we expect that all qualitative (and some quantitative)
  predictions made here will remain unchanged.
\bibitem{commentCancel} In the independent soliton approximation used here
this cancellation is not complete. Although we have not shown this
explicitly, from experience with the analogous calculation for the
sine-Gordon model and physical considerations it is clear that if the
proper multi-soliton solution is used, here too the total $N_s$-soliton
energy $E_{Ns}$ should approach $0$ in the dense soliton limit,
$h\rightarrow h_{c2}$, as the fully incommensurate state, $u_C=h z$ is
approached.
\bibitem{commentNp} Our definition of the pinning number $M_p$ should not be 
confused with the related quantitity $N_p$ made in
Ref.~\onlinecite{Affleck}. Although the two measures of the pinning
influence of the columnar deffect are different, they nearly agree in
the $h\gg h_{c1}$ limit.
\bibitem{commentRGno} Absence of graphical corrections
(renormalization) to $\sqrt{B K}$ is guaranteed by the long-range
nature (nonanalytic form $|q|$) of the corresponding 1D vortex lattice
elasticity. This property results in the exactly vertical
renormalization-group flows of $v(\ell)$ in the $v$ - $\sqrt{B K}$
plane, as for example found by Kane and Fisher\cite{KaneFisher}
\end{references}
\end{document}